\documentclass{sigkddExp}
\usepackage{booktabs} 
\usepackage{algorithmic}
\usepackage{amsmath,amssymb,amsfonts}
\usepackage[table,xcdraw]{xcolor}
\usepackage{epsfig}
\usepackage{multirow}
\usepackage{makecell}
\usepackage{subfigure}
\usepackage{diagbox}
\usepackage{adjustbox}
\usepackage{balance}
\usepackage{tcolorbox}
\usepackage{hyperref}
\usepackage{paralist}
\usepackage{graphicx}
\usepackage{tikz-cd}
\usepackage{soul}
\usepackage{threeparttable}

\newtheorem{myDef}{Definition}

\newtheorem{myPat}{Pattern}

\begin{document}
%

\title{Network-based Fake News Detection:\\A Pattern-driven Approach}
%

\numberofauthors{1}
%


\author{
%
\alignauthor Xinyi Zhou and Reza Zafarani\\
       \affaddr{Data Lab, EECS Department, Syracuse University}\\
       \email{\{zhouxinyi,reza\}@data.syr.edu}
}



\maketitle
\begin{abstract}
Fake news gains has gained significant momentum, strongly motivating the need for fake news research. Many fake news detection approaches have thus been proposed, where most of them heavily rely on news content. However, network-based clues revealed when analyzing news propagation on social networks is an information that has hardly been comprehensively explored or used for fake news detection. We bridge this gap by proposing a network-based pattern-driven fake news detection approach. We aim to study the patterns of fake news in social networks, which refer to the news being spread, spreaders of the news and relationships among the spreaders. Empirical evidence and interpretations on the existence of such patterns are provided based on social psychological theories. These patterns are then represented at various network levels (i.e., node-level, ego-level, triad-level, community-level and the overall network) for being further utilized to detect fake news. The proposed approach enhances the explainability in fake news feature engineering. Experiments conducted on real-world data demonstrate that the proposed approach can outperform the state of the arts.
\end{abstract}

\section{Introduction}
With ``post-truth'' named as the Word of the Year in 2016 by the Oxford Dictionary, discussion around fake news has sparked, especially in the period around the 2016 U.S. presidential election and the U.K. Brexit referendum~\cite{zhou2018fake}. The rise of social media  and its popularity play an indispensable role in this surge of interest. Social media breaks the
physical distance barrier among individuals, and provides rich platforms for users to participate and discuss online news, where the most popular story during the critical months of the 2016 U.S. presidential election campaign (``\textit{Pope Francis Shocks World, Endorses Donald Trump for President, Releases Statement}'', which was fake news) can generate 960,000 shares, reactions, and comments on Facebook~\cite{silverman2016analysis}. 

The situation becomes worse with the existence of an echo chamber effect on social media, where the biased information can be amplified and reinforced~\cite{jamieson2008echo}. Meanwhile, studies have shown that humans can be irrational and vulnerable differentiating between truth and falsehood when overloaded with deceptive information; studies in social psychology and communications have demonstrated that human ability to detect deception is only slightly better than chance - with a mean accuracy of 54\% over 1,000 participants in over 100 experiments~\cite{rubin2010deception}. Various manual fact-checking websites and platforms (e.g., PolitiFact\footnotemark and Snopes\footnotemark) have emerged to serve the public on this matter. Nevertheless, manual fact-checking does not scale well with the volume of newly created information, especially on social media, hence motivating the need for automatic fake news detection.

\footnotetext[1]{\url{https://www.politifact.com/}}
\footnotetext[2]{\url{https://www.snopes.com/}}

Current research on automatic fake news detection heavily relies on news content~\cite{oshikawa2018survey}. These studies have significantly contributed to fake news detection (see ``Related Work'' in Section \ref{sec::review}) while often face multiple challenges. 

First, the traditional approach to detect fake news is to use a \textit{knowledge-based} fact-checking system~\cite{ciampaglia2015computational,shi2016discriminative}. 
The system compares relational knowledge extracted from to-be-verified news content with that stored in a knowledge graph, often a ground truth dataset collected from the Web~\cite{dong2014knowledge,ren2018scalable}. However, the most serious issue by using such system is that it can only detect false news instead of fake news (i.e., intentionally false news)~\cite{reza2019fake}.
Second, another common approach is to use a \textit{style-based} fake news detection system by assuming that fake news exhibits a distinguishable writing style from that of the truth~\cite{zhou2019content}, where malicious entities can disguise the writing style to bypass these linguistic models. 
Recently, neural networks and deep learning techniques have been well developed to detect fake news by incorporating multi-modal or social-network data, e.g., images within news content~\cite{wang2018eann,yang2018ti} and users (news spreaders)~\cite{liu2018early,ruchansky2017csi,zhang2018fake}; nevertheless, these models often face the problems with computational efficiency or interpretability~\cite{zhou2018fake}.

\vspace{0.5em}
\noindent \textbf{Present Work:} Considering that social-network data related to news propagation and spreaders has hardly been comprehensively explored (across network levels) and used in an explainable way for fake news detection,
we propose a network-based pattern-driven fake news detection model, robust against manipulations by malicious entities on news content. To that end, our work aims to utilize patterns in fake news dissemination on social networks, which reveal that compared to the truth, fake news can (i)~spread farther and (ii)~attract more spreaders, where these spreaders are often (iii)~more strongly engaged with the news and (iv)~more densely connected within the network.
Machine learning features representing these patterns are designed at different levels of a network (i.e., node-, ego-, triad-, community-, and network-level), which will be further used within a supervised learning framework to detect fake news. 
Overall, the specific contributions of this paper are as follows: 
\leftmargini=5mm
\leftmarginii=3mm
\begin{enumerate}
\item A network-based pattern-driven approach is proposed, which can detect fake news in an explainable way. Experiments conducted on real-world data demonstrate that the proposed approach can perform comparatively well compared to the state of the art.
\item Fake news patterns in social networks are investigated and summarized, which relate to the  news being spread, spreaders of the news, and relationships among the news spreaders. Empirical studies and social psychological theories are provided to validate and interpret the existence of these patterns;
\item Fake news patterns are represented and quantified across multiple network levels, i.e., node, ego, triad, community, and the overall network level. Experimental results indicate that the proposed approach can perform stably with limited available network information, which makes it suitable for fake news early detection.
\end{enumerate}

The rest of this paper is organized as follows. Section \ref{sec::review} reviews current fake news detection research. Fake news patterns in social networks are summarized and represented in Section \ref{sec::netMeasures}. Experiments are conducted and presented in Section \ref{sec::experiments}. Section \ref{sec::conclusion} concludes the paper.

\section{Related Work} \label{sec::review}
As an emerging topic, the development of fake news detection is in its early stages, where the existing research can be generally grouped into content-based and network-based fake news detection. 

\vspace{0.5em}
\noindent \textbf{Content-based Fake News Detection.} 
Content-based fake news detection investigates news content. One traditional way of detection is based on \textit{knowledge}, often represented as a set of (Subject, Predicate, Object) triples~\cite{dong2014knowledge,nickel2016review}.
Knowledge-based approaches aim to assess news authenticity by comparing the knowledge extracted from to-be-verified news content with true knowledge (i.e., ground truth)~\cite{ciampaglia2015computational,shi2016discriminative}. Such ground truth is generally provided in a knowledge graph such as Knowledge Vault~\cite{dong2014knowledge}, which contains massive manually processed relational knowledge from the open Web. However, the timeliness and completeness of knowledge graphs are still open issues, and importantly, such approaches developed can only detect false news rather than fake news (intentionally false news)~\cite{zhou2018fake}.

Another common way is based on writing \textit{style}, a set of self-defined [non-latent] features well representing news writing style.
Style features can be those capturing content structure
at various language levels such as discourse level by employing
rhetorical structure theory~\cite{rubin2015truth,karimi2019learning}; or those capturing specific attributes in the content such as sentiment and readability~\cite{potthast2017stylometric,perez2017automatic,zhou2019content}, which can be supported by forensic psychological theories such as Undeutsch hypothesis ~\cite{undeutsch1967beurteilung}. Such fundamental theories are a double-edged sword for content-based fake news detection: features inspired can help achieve explainable fake news detection, while some linguistic cues that they reveal might not be applicable for news articles (e.g, non-immediacy)~\cite{zhou2018fake}.

In addition to non-latent features, fake news detection based on latent representation of news content has been well developed recently, where neural networks such as Convolutional Neural Network (CNN)~\cite{wang2018eann} have been utilized to automatically select content features. Nevertheless, these features are often difficult to be comprehended.

While content-based approaches can detect fake news by analyzing news content from various perspectives, auxiliary information revealed in news propagation, e.g., news spreaders, is not considered. In addition, approaches can be sensitive to news content when heavily relying on it, where malicious entities might manipulate the results of detection by disguising their writing styles. Hence, network-based fake news detection has been emerged recently. 

\vspace{0.5em}
\noindent \textbf{Network-based Fake News Detection.} 
Network-based fake news detection utilizes social context information revealed in news propagation. In general, it investigates two types of networks: \textit{homogeneous} and \textit{heterogeneous} networks.

Homogeneous networks contain single type of nodes and edges. A typical example is the stance network, which represents the stance (e.g., \textit{for} or \textit{against}) similarity among news or posts of news. Based on such network, Jin et al. evaluate news credibility by mining the stance correlations within a graph optimization framework~\cite{jin2016news}. 
Another typical example of homogeneous networks is the propagation graph (tree), which presents post-repost relationships for each news article on social media, e.g., tweets and retweets on Twitter~\cite{wu2015false,ma2018rumor}. Using propagation trees, for instance, Vosoughi et al. discover that fake news spreads faster, farther and more broadly than the truth~\cite{vosoughi2018spread}. 

Heterogeneous networks have multiple types of nodes or edges. By exploring relationships among entities such as news articles, publishers, users (spreaders) and user posts, PageRank-like algorithm~\cite{gupta2012evaluating}, matrix/tensor factorization~\cite{gupta2018cimtdetect,shu2019beyond}, and Recurrent Neural Networks (RNN)~\cite{ruchansky2017csi,zhang2018fake} have been developed for fake news detection.

In general, our work is a complement of the current [network-based] studies. Compared to current studies, our work investigates a homogeneous network, the friendship network. To our best knowledge, studying fake news with respect to the friendship network is yet to be explored, which allows one to better understand news spreaders and their social relationships on various network levels. Additionally, we aim to detect fake news in an explainable way - by utilizing its propagation characteristics on social networks, which will be detailed in the next section. 

\begin{table*}[t]
\centering
\caption{Key Notations}
\label{tab::notations}
\begin{tabular}{cl}
\toprule[1pt]
\textbf{Notation} & \textbf{Description} \\ \hline
$\mathcal{F}; \mathcal{T}$ & Fake news events; True news events \\ 
$\mathrm{G}=(\mathrm{V},\mathrm{E})$ & Social (friendship) network \\ 
$\mathrm{G}_X=(\mathrm{V}_X,\mathrm{E}_X)$ & 
$X=\mathcal{F}$: Fake news network; $X=\mathcal{T}$: True news network 
\\ 
$\mathrm{E}_{NS}$ & Relationships from a normal user to a susceptible user \\
$\mathrm{E}_{\triangle > 0};\mathrm{E}_{\triangle = 0};\mathrm{E}_{\triangle < 0}$ & Relationships satisfying $\mathbf{S}(v_i)-\mathbf{S}(v_j)>0;\mathbf{S}(v_i)-\mathbf{S}(v_j)=0;\mathbf{S}(v_i)-\mathbf{S}(v_j)<0$ \\
$\mathrm{V}_X$; $\mathrm{Tr}_X$; $\mathrm{M}_X$ & Nodes (Spreaders); Triads; Communities within $\mathrm{G}_X$ \\ 
$\mathbf{B}(\mathsf{*})$ & $\mathbf{B}=1$ if $\mathsf{*}$ is true; otherwise, $\mathbf{B}=0$ \\ 
$\mathbf{C}(v)$ & Influence (centrality) of user $v$ \\ 
$\mathbf{S}(v)$ & Susceptibility of user $v$ \\ 
$\theta$ & Threshold of user susceptibility, $\mathbf{S}(v)<\theta$ ($\mathbf{S}(v)>\theta$) indicates a normal (susceptible) user \\ 
$\mathbf{T}(v,X)$ & Spreading frequency of user $v$ for news event $X$ \\ 
\bottomrule[1pt]
\end{tabular} 
\end{table*}

\section{Fake News Patterns and Representation in Networks}
\label{sec::netMeasures}
Fake news dissemination in networks exhibits distinguishable patterns from the diffusion of true news. In this section, we summarize these patterns and discuss social psychological theories that can explain the existence of these patterns. In terms of fake news patterns, we demonstrate ways to represent news articles as a set of features across network levels (i.e., node-, triad-, community- and network-level), which can be further utilized to detect fake news within a supervised machine learning framework.

Broadly speaking, fake news patterns involved in this study relate to (1)~the news being spread (Section \ref{subsec:p1} and Section \ref{subsec:p2}),  (2)~spreaders of the news (Section \ref{subsec:p3}), and (3) relationships among the news spreaders (Section \ref{subsec:p4}). Before further elaboration, we first define Fake News Network (FNN) in Definition~\ref{def::fnns}.

\begin{myDef}[Fake News Network, FNN] \label{def::fnns}
Fake \\ News Network (FNN) is a subgraph $\mathrm{G}_{\mathcal{F}}=(\mathrm{V}_{\mathcal{F}}, \mathrm{E}_{\mathcal{F}})$ of the social network $\mathrm{G}=(\mathrm{V},\mathrm{E})$, where $\mathrm{V}_{\mathcal{F}} \in \mathrm{V}$ are the users that have engaged with fake news $\mathcal{F}$, and $\mathrm{E}_{\mathcal{F}} \in \mathrm{E}$ represents the relationships among these users.
\end{myDef} 

\noindent True News Network (TNN) is similarly defined, which is denoted as $\mathrm{G}_{\mathcal{T}}=(\mathrm{V}_{\mathcal{T}}, \mathrm{E}_{\mathcal{T}})$ for a true news event $\mathcal{T}$. 
The key notations in this section are presented in Table \ref{tab::notations}.

\subsection{More-Spreader Pattern}
\label{subsec:p1}

Evidence has been provided that fake news is in general more ``popular'' than true news within the same population of users. For instance, during the critical months of the 2016 U.S. presidential election campaign, top twenty frequently-discussed fake election stories generated 8,711,000 shares, reactions, and comments on Facebook, ironically, greater than the total of 7,367,000 for the top twenty most-discussed election stories posted by nineteen major news medium~\cite{silverman2016analysis}. 
Fake news popularity can be attributed to two reasons. First, as stated by information gap theory~\cite{loewenstein1994psychology}, rather than telling the truth, fake news creators make great efforts to produce an information gap between the news content and individuals' knowledge. Such information gap produces the feeling of deprivation labeled curiosity, which motivates individuals to obtain the missing information to reduce such feeling. Secondly, to greatly influence online users, those who can benefit from fake news often create or recruit malicious accounts (e.g., bots~\cite{shao2018spread}) to spread or discuss the fake content. For example, millions of malicious accounts have participated in 2016 U.S. presidential election online discussions.\footnote{{{https://comprop.oii.ox.ac.uk/research/public-scholarship/resource-for-understanding-political-bots/}}}

News popularity can be characterized in terms of the number of users that spread such news, where Vosoughi et al.~\cite{vosoughi2018spread} have empirically validated that:

\begin{myPat}[More-Spreader Pattern]
More users spread fake news than true news.
\end{myPat} 

To capture the number of news spreaders, we investigate the number and proportion of (\textbf{I}) general (i.e., \textit{non-attributed}) spreaders and (\textbf{II}) specific (i.e., \textit{attributed}) spreaders in news propagation.

\vspace{0.5em}
\noindent \textbf{I. General (Non-Attributed) Spreaders.} In general, the {\textsc{More-Spreaders Pattern}} can be quantified by the number of users involved in spreading each fake or true news story. This number is basically the number of nodes within each FNN and TNN, which we use as a feature. 

\vspace{0.5em}
\noindent \textbf{II. Specific (Attributed) Spreaders.}
Principles like homophily~\cite{mcpherson2001birds} and social validation theory~\cite{cialdini2009influence} suggest that in a social network, users with similar characteristics tend to become connected or form groups and exhibit similar behavior. These observations imply that spreaders of fake (true) news stories may also share some similar attributes; hence, allowing one to distinguish fake news from true news by studying specific users (i.e., with specific attributes) participated in news dissemination. Here we consider (\textit{a}) user susceptibility [to fake news] and (\textit{b}) user influence, both of which are attributes that can be computed with information provided by FNNs and TNNs. 

\vspace{0.5em}
\textit{a. User Susceptibility.}
We investigate user susceptibility to fake news based on (\textit{i}) the number of involvements in the propagation of different fake news and (\textit{ii}) the frequency of such involvements.

\begin{enumerate}
    \item[\textit{i.}] \textit{Number of Involvements.} Susceptibility in terms of involvements is defined as the proportion of fake news among all news that user $v_i$ has participated in spreading, which is denoted as $\mathbf{S}(v_i)$:
\begin{equation}\label{eq::userVirality1}
{{\mathbf{S}(v_i) = \frac{\sum_j \mathbf{B} (v_i \in \mathrm{V}_{\mathcal{F}_j})}{\sum_k \mathbf{B} (v_i \in \mathrm{V}_{\mathcal{T}_k})+\sum_j \mathbf{B} (v_i \in \mathrm{V}_{\mathcal{F}_j})}}},
\end{equation} 
where $\mathbf{B} (v_i \in \mathrm{V}_{X}) = 1$ if $v_i \in \mathrm{V}_{X}$, otherwise $\mathbf{B} (v_i \in \mathrm{V}_{X}) = 0$. $\mathbf{S}(v_i)= 1$ ($\mathbf{S}(v_i)= 0$) indicates that all news stories spread through $v_i$ are fake (true), i.e., $v_i$ is completely susceptible (immune) to fake news. 

\item [\textit{ii.}] \textit{Frequency of Involvements.}
Consider the special case where a user spreads a true news story once and a fake news story multiple times, this user may need to be considered more susceptible than a user who posts each story once. Hence, as an alternative way, we define user susceptibility as the ratio between the spreading frequency of fake news stories and that of all news stories a user has spread. Mathematically, 
\begin{small}
\begin{equation} 
\hspace*{-4mm}
\label{eq::userVirality2}
\mathbf{S}(v_i) =\frac{\sum_j \mathbf{B} (v_i \in \mathrm{V}_{\mathcal{F}_j}) \mathbf{T}(v_i,\mathcal{F}_j)}{\sum_k \mathbf{B} (v_i \in \mathrm{V}_{\mathcal{T}_k}) \mathbf{T}(v_i,\mathcal{T}_k)+\sum_j \mathbf{B} (v_i \in \mathrm{V}_{\mathcal{F}_j}) \mathbf{T}(v_i,\mathcal{F}_j)},
\end{equation} 
\end{small}
where $\mathbf{T}(v_i,X)$ is the number of times that $v_i$ has spread news story $X$. 
\end{enumerate}

Being assigned with a susceptibility score $\mathbf{S}(v_i)$, users can be further labeled as susceptible ($\mathbf{S}(v_i)>\theta$) or normal ($\mathbf{S}(v_i)<\theta$) based on a fixed threshold value $\theta\in [0,1]$. Such labeling  allows us to represent {\textsc{More-Spreaders Pattern}} by recording the (i) number and (ii) proportion of susceptible spreaders (nodes) in each FNN or TNN, as well as the (iii) number and (iv) proportion of normal spreaders within each FNN and TNN. We include (i-iv) as features representing the pattern. Without such labeling one can represent spreaders involved in each FNN or TNN in terms of their mean and median of susceptibility scores, which are also considered into our feature set.

\vspace{0.5em}
\textit{b. User Influence.} An approximation of a node (user) influence can be computed via a centrality score within the network. One can consider the following well-established criteria for computing centrality: (i) [in-, out-] degrees, (ii) [in-, out-] closeness, (iii) betweenness, (iv) PageRank score, (v) hub and authority score, all of which use the connections among nodes to identify their positions within the network. 
We avoid grouping users into influential and non-influential users as many parameters will be introduced (each centrality measure requires a threshold value), which in turn can affect the performance of fake news detection. Therefore, based on each centrality measure, we directly calculate the mean and median user influence within each FNN and TNN, and include both as features.

\subsection{Farther-Distance Pattern}
\label{subsec:p2}

In addition to the number of users that spread news articles, news popularity can be also characterized by how far the news can spread, which leads to the corresponding pattern:
\begin{myPat}[Farther-Distance Pattern]
Fake news \\ spreads farther than true news.
\end{myPat}

This pattern has been observed and validated by Vosoughi et al.~\cite{vosoughi2018spread}; they found that the propagation trees of fake news are generally deeper than that of truth, i.e., an original post referring to a fake news event is often more iteratively forwarded than a true news event. On the other hand, given a news story, how far it spreads can be approximated by computing the shortest ``distance'' between the two most distant spreaders (nodes) within the corresponding FNN or TNN (i.e., network diameter). To represent {\textsc{Farther-Distance Pattern}} and calculate such ``distance'', we investigate (\textbf{I}) {shortest (geodesic) distance} which refers to the paths existing between two nodes, and (\textbf{II}) effective distance which considers the information flow between two nodes~\cite{brockmann2013hidden}.

\vspace{0.5em}
\noindent \textbf{I. Geodesic Distance.} Based on geodesic distance, the diameter of each FNN and TNN is equivalent to the shortest path length between the two most distant spreaders within the network.

\vspace{0.5em}
\noindent \textbf{II. Effective Distance.} Besides conventional shortest distance, we introduce effective distance to help assess the network diameter, which was initially proposed by Brockmann and Helbing~\cite{brockmann2013hidden}. The initial binary (unweighted) FNNs and TNNs is hence transformed into weighted networks, where the weights are determined by the volume of information flow among nodes. Given a network, the effective distance among nodes is defined as follows. 

\begin{myDef}[Effective Distance] 
\label{def::effectiveDistance}
Given a network $\mathrm{G}$, we assume $\mathrm{F}$ denotes the flow matrix whose entities $\mathrm{F}_{ij}$ represent the volume of information flow from node $i$ to node $j$. Based on the flow matrix, the effective distance $d_{\mathsf{Eff}}(i,j)$ from node $i$ to node $j$ is defined as
\begin{equation} 
d_{\mathsf{Eff}}(i,j) = 1-\log  \frac{\mathrm{F}_{ij}}{\sum_l \mathrm{F}_{lj}},
\end{equation}
where $d_{\mathsf{Eff}}(i,j)$ satisfies $d_{\mathsf{Eff}(i,j)}\geq1$.
\end{myDef} 

Information flow has been defined differently in various networks. For instance, it can be the passenger flux in global mobility networks or the transport flow in transportation networks~\cite{brockmann2013hidden}.
In FNNs and TNNs it is the news flow among nodes (users) in the network which can be defined as (i) the total number of news stories both users have spread, i.e., $\mathrm{F}_{ij}=\sum_X \mathbf{B}(e_{ij} \in \mathrm{E}_X)$, or (ii) the overall number of times both users have at least spread the same news stories, i.e., $\mathrm{F}_{ij}=\sum_X \mathbf{B}(e_{ij} \in \mathrm{E}_X) \times \min \{\mathbf{T}(u_i, X), \mathbf{T}(u_j, X) \}$. 
The diameter of each FNN and TNN based on effective distance is then equivalent to the minimum [sum of] effective distance between the two most distant spreaders within the network. We include diameters computed using geodesic and effective distances as features representing {\textsc{Farther-Distance Pattern}}.


\subsection{Stronger-Engagement Pattern}
\label{subsec:p3}

The statistics in \cite{silverman2016analysis} have revealed that fake news stories can  engage users more compared to true news stories. 
Note that a user may decide to engage with a fake news story (e.g., post it)  more than one time, such ``more engagements'' can be attributed to the number of users engaging with fake news, which has been summarized as {\textsc{More-Spreader Pattern}} investigated in Section \ref{subsec:p1}, and/or to the number of times each user engages with a fake news story, leading to the following pattern:

\begin{myPat}[Stronger-Engagement Pattern]
~\\ Spreaders engage more strongly with fake news than with true news.
\end{myPat} 

To quantify the ``engagements'' of users for each news story, one can concentrate on (\textbf{I}) group level engagements, i.e., the engagements of all spreaders, and (\textbf{II}) individual level engagements, i.e., the engagements of a single spreader.

\vspace{0.5em}
\noindent \textbf{I. Group Engagements.} On a group level, quantifying spreader engagements for a certain news story can be equivalent to counting the total number of times that the news story has been spread. With specific user attributes (susceptible or normal), such engagements can be further quantified as the (i) number or (ii) proportion of times that the news story has been spread by susceptible users, as well as (iii) number or (iv) proportion of times that the news story has been spread by normal users.

\vspace{0.5em}
\noindent \textbf{II. Individual Engagements.} Individual engagements of a news story can be evaluated by the average spreading frequencies of (susceptible, normal, all) users who have participated in the news propagation.
In this case, the impact of the number of such news spreaders (i.e., {\textsc{More-Spreaders Pattern}}) is divided and removed. \vspace{1mm}

All above ways of representing fake news patterns are on the level of nodes, e.g., individual engagement, and the whole network, e.g., network diameter. Next we will specify how to represent {\textsc{Denser-Networks Pattern}} for fake news detection, which will be represented at different network levels: ego, triad and community.


\subsection{Denser-Network Pattern}
\label{subsec:p4}

Research in social psychology such as  homophily~\cite{mcpherson2001birds} and social validation theory~\cite{cialdini2009influence} suggests that connected users in social networks share similar attributes, interests and behaviors, e.g., sharing the same news article. On the other hand, malicious users often form cohesive groups, taking collective action that are more purposeful than normal users~\cite{mukherjee2013spotting,xie2012review}. These fundamental theories suggest the possibility to assume that fake and true news articles can be distinguished by the relationships among their corresponding spreaders, which can be summarized as: 

\begin{myPat}[Denser-Network Pattern]
Fake news spreaders form denser networks compared to truth spreaders.
\end{myPat}

To capture the ``density'' of connections among news spreaders, we analyze news networks at different levels: (\textbf{I}) ego, (\textbf{II}) triad and (\textbf{III}) community levels.

\vspace{0.5em}
\noindent \textbf{I. Ego Level.} At the ego level, to compute density of networks formed by users that have engaged with a certain news story, we look at the numbers and proportions of connections that these users (spreaders) have (\textit{i}) generally formed with other spreaders, and (\textit{ii}) specifically with other normal or susceptible spreaders.

\vspace{0.3em}
\noindent \textit{i. General Ego Relations.} We include as a feature the total number of ego relationships among spreaders for each news story, i.e., the number of edges within each FNN and TNN ($|\mathrm{E}_X|$). To eliminate the impact of the number of news spreaders (i.e., {\textsc{More-Spreaders Pattern}}), for each FNN or TNN $\mathrm{G}_X$ we also record $|\mathrm{E}_X|/|\mathrm{V}_X|$ and $|\mathrm{E}_X|/\binom{|\mathrm{V}_X|}{2}$, which calculate the average number of ego relationships per spreader and network density, respectively. Here, ${|\mathrm{V}_X|}$ is the number of spreaders (nodes) in $\mathrm{G}_X$ and $\binom{|\mathrm{V}_X|}{2}$ is the number of edges within a fully connected version of $\mathrm{G}_X$.

\vspace{0.3em}
\noindent \textit{ii. Specific Ego Relations.}
Labeling users as susceptible or normal allows one to group all directed ego relationships into four subsets: (1)~$\mathrm{E}_{NN}$ containing relationships from a normal user to a normal user, (2)~$\mathrm{E}_{NS}$ containing relationships from a normal user to a susceptible one, (3)~$\mathrm{E}_{SN}$ containing relationships from a susceptible user to a normal one, (4)~$\mathrm{E}_{SS}$ containing relationships from a susceptible user to a susceptible one. We include the number and proportion of each type of edges within a FNN or TNN as features being used for fake news detection. In addition, each edge $e_{ij}$ can be also classified into one of the following set: (1) $\mathrm{E}_{\triangle>0}$ if $\triangle = \mathbf{S}(v_i)-\mathbf{S}(v_j)>0$, (2) $\mathrm{E}_{\triangle=0}$ if $\mathbf{S}(v_i)-\mathbf{S}(v_j)=0$, (3) $\mathrm{E}_{\triangle<0}$ if $\mathbf{S}(v_i)-\mathbf{S}(v_j)<0$ which does not require partitioning users into susceptible or normal ones. We also include as features the number and proportion of each above type of edges within a FNN or TNN. 

\vspace{0.5em}
\noindent \textbf{II. Triad Level.} Triads (a set of three connected users) are the most basic subgraphs of networks. Similar to our study at the ego level, we investigate (i) general triads and (ii) specific triads formed between [susceptible and normal] users within networks.

\vspace{0.3em}
\noindent \textit{i. General Triads.} One simple way to represent the \textsc{Denser-Network pattern} is to directly count the total number of triads $|\mathrm{Tr}_X|$ within a $\mathrm{G}_X$. Similarly, to control for {\textsc{More-Spreaders Pattern}}, we also include as features the value of $|{\mathrm{Tr}_X}|/|\mathrm{V}_X|$ and $|\mathrm{Tr}_X|/\binom{|\mathrm{V}_X|}{3}$ where $\binom{|\mathrm{V}_X|}{3}$ is the number of triads within a fully connected version of $\mathrm{G}_X$.

\begin{figure}[t]
    \centering
    \includegraphics[width=0.45\textwidth]{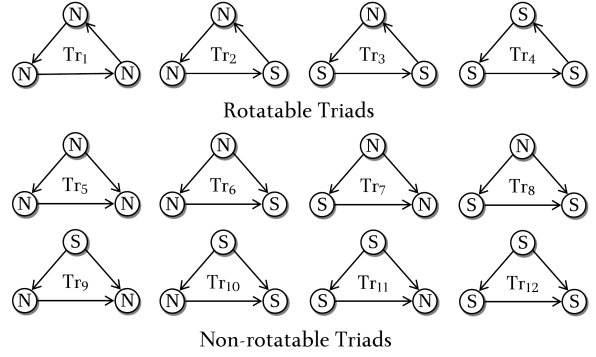}
    \vspace{-2mm}
    \caption{Specific Triads. $N$ indicates normal users and $S$ indicates susceptible users. $A\rightarrow B$ denotes  $A$ follows $B$.}
    \vspace{-4mm}
    \label{fig::labeledTriads}
\end{figure}

\vspace{0.3em}
\noindent \textit{ii. Specific Triads.} Regarding each user as either a susceptible or normal user, we can have twelve different triads to be further explored (shown in Figure \ref{fig::labeledTriads}). We include as features the number and proportion of every type of triads within each FNN and TNN. 

\vspace{0.5em}
\noindent \textbf{III. Community Level.}
In networks, a community structure refers to the occurrence of groups of nodes in a network that are more densely connected internally than with the rest of the network. Therefore, the number and proportion of communities within each FNN and TNN can be used to represent {\textsc{Denser-Networks Pattern}} and, broadly speaking, should be negatively correlated to the network density. 

As features, we include the number of communities $|\mathrm{M}_X|$ within each FNN and TNN, and the proportion of communities (assuming in the worst case each node is its own community) which removes the impact of the number of news spreaders, i.e., the value of $|\mathrm{M}_X| / |\mathrm{V}_X|$. Note that $|\mathrm{M}_X|$ can be obtained either from (i) global or (ii) local perspective. From a global perspective, communities that nodes (spreaders) belong to within a FNN or TNN, as a subgraph of the social network, are based on the structure of the overall social network. From a local perspective, communities can be detected within a FNN or TNN. We include counts and proportion features for both types of communities.


\begin{table*}
\renewcommand\arraystretch{1.5}
\caption{Network-based Pattern-driven Feature Set for Fake News Detection}
\label{tab::features}
\begin{adjustbox}{width = \textwidth}
\begin{tabular}{|c|c|p{11cm}|c|}
\toprule[1pt]
\textbf{Pattern} & \textbf{No.} & {\textbf{Feature(s)}} & \textbf{Formulation(s)} \\ \midrule 
\multirow{12}{*}{\renewcommand\arraystretch{1}\rotatebox{90}{\begin{tabular}[c]{@{}c@{}} \small \textsc{More-}\\ \textsc{Spreaders} \\ \textsc{Pattern}  \end{tabular}}} & 1 & {\# News Spreaders} & $|\mathrm{V}_X|$\\ \cline{2-4} 
& \multirow{5}{*}{2-9} 
& \# Normal Spreaders, where user susceptibility, a.k.a., $\mathbf{S}(v)$, is based on Equation (\ref{eq::userVirality1}) (\#news) or Equation (\ref{eq::userVirality2}) (frequency) 
& \multirow{2}{*}{$|\sum_{v_j \in \mathrm{V}_X}\mathbf{B}(\mathbf{S}(v_j) \leq \theta)|$} \\ \cline{3-4} 

&  
& \# Susceptible Spreaders, where $\mathbf{S}(v)$ is based on \#news or frequency & $|\sum_{v_j \in \mathrm{V}_X}\mathbf{B}(\mathbf{S}(v_j) \geq \theta)|$ \\ \cline{3-4} 

&  & \% Normal Spreaders, where $\mathbf{S}(v)$ is based on \#news or frequency  & $\frac{|\sum_{v_j \in \mathrm{V}_X}\mathbf{B}(\mathbf{S}(v_j) \leq \theta)|}{|\mathrm{V}_X|}$ \\ \cline{3-4} 

&  & \% Susceptible Spreaders, where $\mathbf{S}(v)$ is based on \#news or frequency  & $\frac{|\sum_{v_j \in \mathrm{V}_X}\mathbf{B}(\mathbf{S}(v_j) \geq \theta)|}{|\mathrm{V}_X|}$ \\ \cline{2-4}  

& \multirow{2}{*}{10-13} & Average Spreader Susceptibility, where $\mathbf{S}(v)$ is based on \#news or frequency & $ \frac{\sum_{v_j \in \mathrm{V}_X} \mathbf{S}(v_j)}{|\mathrm{V}_X|}$ \\ \cline{3-4}

& & Median Spreader Susceptibility, where $\mathbf{S}(v)$ is based on \#news or frequency & \makecell{$\mathbf{P}(\mathbf{S}(v_j)\leq\text{MSS})=0.5$ \\ for $v_j \in \mathrm{V}_X$}\\ \cline{2-4}

& \multirow{3}{*}{14-29} 
& Average Spreader Influence, where influence is based on (in-, out-) degree, (in-, out-) closeness, betweenness, PageRank score, hub and authority score 
& \multirow{2}{*}{$\frac{\sum_{v_j \in \mathrm{V}_X} \mathbf{C}(v_j)}{|\mathrm{V}_X|}$} \\ \cline{3-4} 

& 
& Median Spreader Influence, where influence is based on (in-, out-) degree, (in-, out-) closeness, betweenness, PageRank score, hub and authority score 
& \multirow{2}{*}{\makecell{$\mathbf{P}(\mathbf{C}(v_j)\leq\text{MSI})=0.5$ \\ for $v_j \in \mathrm{V}_X$}} \\ \midrule

\multirow{3}{*}{\rotatebox{90}{\renewcommand\arraystretch{1}\begin{tabular}[c]{@{}c@{}}\small \textsc{Farther-}\\ \textsc{Distance} \\ \textsc{Pattern} \end{tabular}}} 
& 30-32 
& Maximum, Average, and Median Geodesic Distance & - \\ \cline{2-4} 

& \multirow{2}{*}{33-38} 
& Maximum, Average, and Median Effective Distance, information flow is based on \#news and frequency & \multirow{2}{*}{See Definition \ref{def::effectiveDistance}} \\ \midrule

\multirow{10}{*}{\rotatebox{90}{\renewcommand\arraystretch{1}\begin{tabular}[c]{@{}c@{}}\small \textsc{{Stronger-}}\\ \textsc{Engagement} \\ \textsc{{Pattern}} \end{tabular}}} 
& 39 
& \# User Engagements & $\sum_{v_j \in \mathrm{V}_X} \mathbf{T}(v_j,X)$
 \\ \cline{2-4}

& \multirow{5}{*}{40-47}
& \# Normal User Engagements, where $\mathbf{S}(v)$ is based on \#news or frequency & $\sum_{v_j \in \mathrm{V}_X; \mathbf{S}(v_j)\leq \theta} \mathbf{T}(v_j,X)$  \\ \cline{3-4}

& & \# Susceptible User Engagements, where $\mathbf{S}(v)$ is based on \#news or frequency & $\sum_{v_j \in \mathrm{V}_X; \mathbf{S}(v_j)\geq \theta} \mathbf{T}(v_j,X)$  \\ \cline{3-4}

& & \% Normal User Engagements, where $\mathbf{S}(v)$ is based on \#news or frequency & $\frac{\sum_{v_j \in \mathrm{V}_X; \mathbf{S}(v_j)\leq \theta} \mathbf{T}(v_j,X)}{\sum_{v_j \in \mathrm{V}_X} \mathbf{T}(v_j,X)}$\\ \cline{3-4}

& & \% Susceptible User Engagements, where $\mathbf{S}(v)$ is based on \#news or frequency & $\frac{\sum_{v_j \in \mathrm{V}_X; \mathbf{S}(v_j)\geq \theta} \mathbf{T}(v_j,X)}{\sum_{v_j \in \mathrm{V}_X} \mathbf{T}(v_j,X)}$  \\ \cline{2-4} 

& 48 & Average User Engagements & $\frac{\sum_{v_j \in \mathrm{V}_X} \mathbf{T}(v_j,X)}{|\mathrm{V}_X|}$
 \\ \cline{2-4} 

& \multirow{3}{*}{49-52} & Avg. Normal User Engagements, where $\mathbf{S}(v)$ is based on \#news or frequency & $\frac{\sum_{v_j \in \mathrm{V}_X; \mathbf{S}(v_j)\leq \theta} \mathbf{T}(v_j,X)}{|\sum_{v_j \in \mathrm{V}_X}\mathbf{B}(\mathbf{S}(v_j) \leq \theta)|}$ \\ \cline{3-4}

& & Avg. Susceptible User Engagements, $\mathbf{S}(v)$ is based on \#news or frequency & $\frac{\sum_{v_j \in \mathrm{V}_X; \mathbf{S}(v_j)\geq \theta} \mathbf{T}(v_j,X)}{|\sum_{v_j \in \mathrm{V}_X}\mathbf{B}(\mathbf{S}(v_j) \geq \theta)|}$  \\ \midrule 

\multirow{19}{*}{\renewcommand\arraystretch{1}\rotatebox{90}{\begin{tabular}[c]{@{}c@{}}\small \textsc{Denser-}\\ \textsc{Networks} \\ \textsc{Pattern} \end{tabular}}} & 53 & {\# Relationships among Spreaders} & $|\mathrm{E}_X|$  \\ \cline{2-4} 

& 54 & {Average \# Relationships of Spreaders} & $|\mathrm{E}_X|/|\mathrm{V}_X|$  \\ \cline{2-4} 

& 55 & {Ego Density} & $|\mathrm{E}_X|/\binom{|\mathrm{V}_X|}{2}$ \\ \cline{2-4} 

& \multirow{4}{*}{56-71} & \#/\% $N \rightarrow N$, where $\mathbf{S}(v)$ is based on \#news and frequency & $|\mathrm{E}_{NN} \cap  \mathrm{E}_{X}|$; $\frac{|\mathrm{E}_{NN} \cap \mathrm{E}_{X}|}{|\mathrm{E}_X|}$ \\ \cline{3-4} 
&  & \#/\% $N \rightarrow S$, where $\mathbf{S}(v)$ is based on \#news and frequency & $|\mathrm{E}_{NS} \cap  \mathrm{E}_{X}|$; $\frac{|\mathrm{E}_{NS} \cap \mathrm{E}_{X}|}{|\mathrm{E}_X|}$ \\ \cline{3-4} 

&  & \#/\% $S \rightarrow N$, where $\mathbf{S}(v)$ is based on \#news and frequency &  $|\mathrm{E}_{SN} \cap  \mathrm{E}_{X}|$; $\frac{|\mathrm{E}_{SN} \cap \mathrm{E}_{X}|}{|\mathrm{E}_X|}$ \\ \cline{3-4} 

&  & \#/\% $S \rightarrow S$, where $\mathbf{S}(v)$ is based on \#news and frequency &  $|\mathrm{E}_{SS} \cap  \mathrm{E}_{X}|$; $\frac{|\mathrm{E}_{SS} \cap \mathrm{E}_{X}|}{|\mathrm{E}_X|}$ \\ \cline{2-4} 

& \multirow{3}{*}{72-83} & \#/\% $\mathbf{S}(v_i)>\mathbf{S}(v_j)$, where $\mathbf{S}(v)$ is based on \#news and frequency & $|\mathrm{E}_{\triangle>0} \cap  \mathrm{E}_{X}|$; $\frac{|\mathrm{E}_{\triangle>0} \cap  \mathrm{E}_{X}|}{|\mathrm{E}_X|}$ \\ \cline{3-4} 

&  & \#/\% $\mathbf{S}(v_i)=\mathbf{S}(v_j)$, where $\mathbf{S}(v)$ is based on \#news and frequency &  $|\mathrm{E}_{\triangle=0} \cap  \mathrm{E}_{X}|$; $\frac{|\mathrm{E}_{\triangle=0} \cap  \mathrm{E}_{X}|}{|\mathrm{E}_X|}$ \\ \cline{3-4} 

&  & \#/\% $\mathbf{S}(v_i)<\mathbf{S}(v_j)$, where $\mathbf{S}(v)$ is based on \#news and frequency &  $|\mathrm{E}_{\triangle<0} \cap  \mathrm{E}_{X}|$; $\frac{|\mathrm{E}_{\triangle<0} \cap  \mathrm{E}_{X}|}{|\mathrm{E}_X|}$  \\ \cline{2-4} 

& 84 & {\# Triads} & $|\mathrm{Tr}_X|$ \\ \cline{2-4} 

& 85 & {Average \# Triads of Spreaders} & $|{\mathrm{Tr}_X}|/|\mathrm{V}_X|$ \\ \cline{2-4} 

& 86 & {Triad Density} & $|\mathrm{Tr}_X|/\binom{|\mathrm{V}_X|}{3}$  \\ \cline{2-4} 

& 87-110 & \# $\mathrm{Tr}_{1}$ to $\mathrm{Tr}_{12}$ (see Figure \ref{fig::labeledTriads}), where $\mathbf{S}(v)$ is based on \#news and frequency
& \multirow{2}{*}{\begin{tabular}[c]{@{}c@{}} $|\mathrm{Tr}_{k} \cap  \mathrm{Tr}_{X}|$; $\frac{|\mathrm{Tr}_{k} \cap  \mathrm{Tr}_{X}|}{|\mathrm{Tr}_{X}|}$; \\ $k=1,2,\cdots,12$ \end{tabular}} \\ \cline{2-3} 

& 111-134 & \% $\mathrm{Tr}_{1}$ to $\mathrm{Tr}_{12}$, where user susceptibility is based on \# news and frequency & \\ \cline{2-4} 

& {135-136} & \# Communities (from global and local perspective, see Section~\ref{subsec:p4} for details) & $|\mathrm{M}_X|$ \\ \cline{2-4} 

& {137-138} & Community Density (from global and local perspective) & $|\mathrm{M}_X|/|\mathrm{V}_X|$  \\ \bottomrule[1pt]
\end{tabular}
\end{adjustbox}
\end{table*}

\vspace{0.5em}
\noindent \textbf{Integrated Representation of Patterns.}
To represent each fake news pattern, we have used network information such as network diameters, the number of news spreaders (size), and the number of relationships among the spreaders (density). Networks with various diameters, sizes and densities exhibit various overall structures. Hence, the overall network structure can be regarded as the integrated representation of all related patterns. On the other hand, including such ``structure'' features to detect fake news helps to evaluate if the fake news patterns and their representations defined in this section have well captured the difference of dissemination between fake news and the truth. To quantify such ``structure'', one can compare the similarities among FNNs and TNNs, where graph kernel and graph embedding~\cite{shervashidze2011weisfeiler,wu2015false} methods can be useful. Here, we consider FNNs and TNNs as \textit{labeled graphs} for further comparison, where node labels can be either (i) user identities or (ii) user attributes (susceptible or normal).

\vspace{0.5em}
Overall, Table \ref{tab::features} presents all features defined and involved in our work to detect fake news, and their corresponding formulations for reproducibility.

\section{Experiments}
\label{sec::experiments}

Fake news patterns in networks have been specified as well as how they can be represented as a set of quantifiable and meaningful features. 
In this section, various experiments are conducted to verify the effectiveness of the proposed approach in detecting fake news. We first present the experimental setup in Section \ref{subsec::setup}, followed by the evaluations of experimental results in Section \ref{subsec::performance}.

\begin{table}[t]
\centering
\caption{Data Statistics}
\label{tab::datasets}
\begin{tabular}{lrr}
\toprule[1pt]
\multicolumn{1}{l}{\textbf{Data}} & \multicolumn{1}{c}{\textbf{PolitiFact}} & \multicolumn{1}{c}{\textbf{BuzzFeed}} \\ \hline
\# Users & 23,865 & 15,257 \\ 
\# News--Users & 32,791 & 22,779 \\ 
\# Users--Users & 574,744 & 634,750 \\
\# News Stories & 240 & 182 \\ 
\# True News & 120 & 91 \\ 
\# Fake News & 120 & 91 \\ 
\# Triads & 6,972,189 & 6,885,951 \\ 
\# Communities & 163 & 46 \\ \bottomrule[1pt] 
\end{tabular}
\end{table}

\subsection{Experimental Setup}
\label{subsec::setup}

We detail data used in experiments in Section \ref{subsubsec::dataset}, followed by how data is prepared for experiments in Section \ref{subsubsec::dataPreparation}, and the baselines which the proposed approach is compared with in Section \ref{subsubsec::baselines}. 

\subsubsection{Datasets}
\label{subsubsec::dataset}
Our experiments are conducted on two public benchmark datasets of fake news detection~\cite{shu2019beyond}.
News articles in these datasets are collected from PolitiFact and BuzzFeed, respectively. Ground truth labels (\textit{true} or \textit{fake}) of news articles in both datasets are provided by fact-checking experts.
In addition to (i) news content and labels, both datasets also provide information on (ii) social network of Twitter which contains Twitter users and their following relationships, i.e., user-user relationships, and (iii) how the news has propagated (tweeted/re-tweeted) by users, i.e., news-user relationships. Based on the original datasets, we further identify triads and communities in the social network. Communities are detected using Louvain algorithm, a fast and widely-accepted modularity-based community detection algorithm~\cite{blondel2008fast}. Statistics of two datasets are shown in Table~\ref{tab::datasets}.

\subsubsection{Data Preparation}
\label{subsubsec::dataPreparation}

Following dataset collection, feature values are computed for both datasets, which will be utilized in a supervised learning framework for fake news detection. However, an extra step is necessary to take when computing user susceptibility scores. In Section \ref{sec::netMeasures}, two ways are defined for determining user susceptibility [to fake news] (see Equation (\ref{eq::userVirality1}) and (\ref{eq::userVirality2}), respectively). Both ways rely on the historical information of users on how they previously engaged in fake news dissemination, where the news labels (\textit{true} or \textit{fake}) are necessary in the calculation. To avoid \textit{information leakage} (i.e., features having an unfair prior knowledge of labels), when dividing a dataset into the training and testing one, we dynamically calculate user susceptibility by using the historical information provided in training dataset, rather than the whole dataset. For users with no historical information in training dataset, we treat their susceptibility as the threshold value, indicating that their susceptibility to fake news is unknown.

\subsubsection{Baselines}
\label{subsubsec::baselines} 

The performance of the proposed method is compared with several benchmark fake news detection methods on the same datasets. These methods include  (1) content-based (linguistic) models, which rely on non-latent (\cite{perez2017automatic,zhou2019content}) or latent representation (\cite{mikolov2013efficient,le2014distributed}) of news content, (2) network-based models (\cite{castillo2011information}), which investigate information revealed in news propagation, and hybrid models (\cite{shu2019beyond}), which utilize both content and network information to detect fake news.

\vspace{0.3em}
\noindent \textbf{I. P\'erez-Rosas et al.~\cite{perez2017automatic}} propose a comprehensive linguistic model for fake news detection, involving the following features:  (i) $n$-grams (i.e., uni-grams and bi-grams) and (ii) CFGs based on TF-IDF encoding; (iii) word and phrase proportions referring to all categories provided by LIWC; and (iv) readability. Features are computed and used to predict fake news within a supervised machine learning framework.

\vspace{0.3em}
\noindent \textbf{II. Zhou et al.~\cite{zhou2019content}}. In our previous study, forensic psychological theories are studied and used to detect fake news in a supervised learning framework, which provide the evidence of distinguishing fake news in content style from the truth. Such content style is captured by the frequency of (i) lexicons relying on Bag-Of-Words (BOW) model, (ii) Part-Of-Speech (POS) tags and Context Free Grammers (CFGs) at syntax-level, (iii) Rhetorical Relationships (RRs) at discourse-level, and by assessing a set of theory-driven (iv) DisInformation-related Attributes (DIAs) and (v) ClickBait-related Attributes (CBAs) at semantic-level. 

\vspace{0.3em}
\noindent \textbf{III. Castillo et al.~\cite{castillo2011information}} design features that exploit information from user profiles, tweets and propagation trees to evaluate news credibility within a supervised learning framework. Specifically, these features are based on (i) quantity, sentiment, hash-tag and URL information from user tweets, (ii) user profiles such as registration age, (iii) news topics through mining tweets of users, and (iv) propagation trees (e.g., the number of propagation trees for each news topic).

\vspace{0.3em}
\noindent \textbf{IV. Shu et al.~\cite{shu2019beyond}} detect fake news by exploring and embedding the relationships among news articles, publishers and spreaders on social media. Such embedding involves (i) news content by using non-negative matrix factorization, (ii) users on social media, (iii) news-user relationships (i.e., user engagements in spreading news articles), and (iv) news-publisher relationships (i.e., publisher engagements in publishing news articles). Fake news detection is then conducted within a semi-supervised machine learning framework.

\vspace{0.3em}
Additionally, fake news detection based on latent representation of news articles is also investigated in comparative studies, where we consider as baselines supervised classifiers with  features being \textbf{(V) \textsc{Word2Vec}}~\cite{mikolov2013efficient} and  \textbf{(VI) \textsc{Doc2Vec}}~\cite{le2014distributed} embedding of news articles.

\begin{figure}[t]
    \centering
    \includegraphics[width=0.48\textwidth]{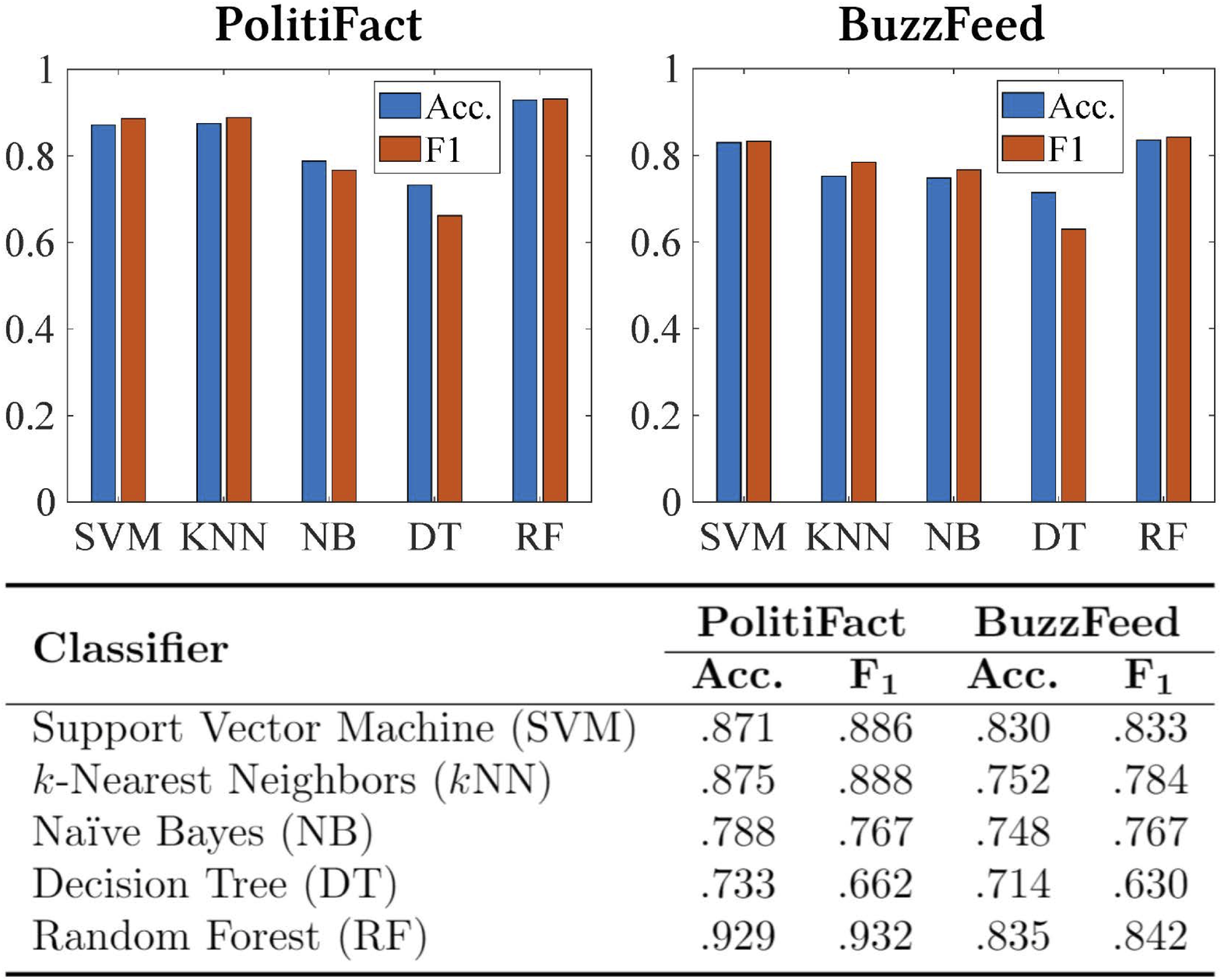}
    \caption{General Performance of Fake News Detection by Using Different Classifiers, where random forests perform best among all on both datasets.}
    \label{fig:classifiers}
\end{figure}

\begin{table}[t]
\centering
\caption{General Performance of Fake News Detection Methods. The proposed network-based approach can perform relatively well compared to the content-based (\protect\cite{perez2017automatic,zhou2019content,mikolov2013efficient,le2014distributed}) and network-based approaches (\protect\cite{castillo2011information}) among baselines. Compared to the hybrid one (\protect\cite{shu2019beyond}), the proposed approach can be comparable with it and can outperform it when introducing the linguistic features in the proposed approach (``Our Approach + \protect\cite{zhou2019content}'').}
\label{tab::comparativeStudies}
\begin{tabular}{lcc|cc}
\toprule[1pt]
\multirow{2}{*}{\textbf{Method}} & \multicolumn{2}{c|}{\textbf{PolitiFact}} & \multicolumn{2}{c}{\textbf{BuzzFeed}} \\ \cline{2-5}
 & \textbf{Acc.}  & $\mathbf{F_1}$ & \textbf{Acc.} & $\mathbf{F_1}$ \\ \hline
{P\'erez-Rosas et al.~\cite{perez2017automatic}} & .811  & .811 & .755  & .757 \\
{Zhou et al.~\cite{zhou2019content}} & .865 & .865 & .855 & .856 \\
$\bullet$ BOWs & .856 & .858 & .823 & .823 \\
$\bullet$ POS Tags & .755 & .755 & .745 & .745 \\
$\bullet$ CFGs & .877 & .877 & .778 & .778 \\
$\bullet$ DIAs & .729 & .735 & .667 & .647 \\
$\bullet$ CBAs & .604 & .612 & .638 & .628 \\
$\bullet$ RRs & .621 & .621 & .658 & .658 \\
{\textsc{Word2Vec}-based~\cite{mikolov2013efficient}} & .688 & .667 & .703 & .718 \\
{\textsc{Doc2Vec}-based~\cite{le2014distributed}} & .698 & .698 & .615 & .615 \\ 
{Castillo et al.~\cite{castillo2011information}} & .794 & .822 & .789 & .794 \\
{Shu et al.~\cite{shu2019beyond}} & .878 & .880 & .864 & .870 \\ \hline
{Our Approach} & .929 & .932 & .835 & .842 \\
{Our Approach + \cite{zhou2019content}} & .933 & .939 & .865 & .884 \\
\bottomrule[1pt]
\end{tabular}
\end{table}

\begin{table*}[t]
\centering
\caption{Pattern Performance in Fake News Detection.
\textsc{More-Spreader Pattern} and \textsc{Stronger-Engagement Pattern} perform best compared to the others when being separately utilized to detect fake news.
When combining different patterns, their performance is in general better than when separately using them, and than when using network similarity, as a mix of patterns from a higher view.}
\label{tab::patternPerformance}
\begin{tabular}{lcc|cc}
\toprule[1pt]
\multirow{2}{*}{\textbf{Pattern(s)}} & \multicolumn{2}{c}{\textbf{PolitiFact}} & \multicolumn{2}{c}{\textbf{BuzzFeed}} \\ \cline{2-5}
 & \textbf{Accuracy} & \textbf{$\mathbf{F_1}$ Score} & \textbf{Accuracy} & \textbf{$\mathbf{F_1}$ Score} \\ \hline
\textsc{More-Spreaders} & .891 & .901 & .808 & .817 \\
\textsc{Farther-Distance} & .639 & .587 & .678 & .698 \\
\textsc{Stronger-Engagement} & .898 & .898 & .807 & .808 \\
\textsc{Denser-Networks} & .746 & .718 & .687 & .704 \\ \hline
\textsc{More-Spreaders} + \textsc{Farther-Distance} & .846 & .803 & .824 & .824 \\
\textsc{More-Spreaders} + \textsc{Stronger-Engagement} & .879 & .864 & .830 & .847 \\
\textsc{More-Spreaders} + \textsc{Denser-Networks} & .919 & .919 & .770 & .796 \\
\textsc{Farther-Distance} + \textsc{Stronger-Engagement} & .917 & .923 & .814 & .824 \\
\textsc{Farther-Distance} + \textsc{Denser-Networks} & .742 & .710 & .786 & .798 \\
\textsc{Stronger-Engagement} + \textsc{Denser-Networks} & .921 & .926 & .829 & .840 \\ \hline
All Patterns - \textsc{Denser-Networks} & .908 & .916 & .814 & .819 \\
All Patterns - \textsc{Stronger-Engagement} & .929 & .928 & .819 & .815 \\
All Patterns - \textsc{Farther-Distance} & .913 & .914 & .780 & .759 \\
All Patterns - \textsc{More-Spreaders} & .879 & .871 & .802 & .803 \\ \hline
All Patterns & .929 & .928 & .828 & .823 \\
Network Similarity (Mix of Patterns)\footnotemark & .808 & .770 & .671 & .689 \\ \hline
All Patterns + Network Similarity & .929 & .932 & .835 & .842 \\ \bottomrule[1pt]
\end{tabular}
\end{table*}

\begin{table*}[t]
\centering
\caption{Top 20 Important Features}
\label{tab::top20}
\begin{threeparttable}
\begin{tabular}{cll}
\toprule[1pt]
\textbf{Rank} & \textbf{PolitiFact} & \textbf{BuzzFeed} \\ \hline
1 & \cellcolor[HTML]{ECF4FF}{\color[HTML]{000000} Avg. Spreader susceptibility (\#News)} & \cellcolor[HTML]{ECF4FF}Median Spreader susceptibility (Frequncy) \\
2 & \cellcolor[HTML]{ECF4FF}{\color[HTML]{000000} Avg. Spreader susceptibility (Frequncy)} & \cellcolor[HTML]{ECF4FF}Median Spreader susceptibility (Frequncy) \\
3 & \cellcolor[HTML]{ECF4FF}{\color[HTML]{000000} Median Spreader susceptibility (Frequncy)} & \cellcolor[HTML]{ECF4FF}Avg. Spreader susceptibility (\#News) \\
4 & \cellcolor[HTML]{ECF4FF}{\color[HTML]{000000} Median Spreader susceptibility (Frequncy)} & \cellcolor[HTML]{ECF4FF}Avg. Spreader susceptibility (Frequncy) \\
5 & \cellcolor[HTML]{E5FFD6}{\color[HTML]{000000} Avg. Normal User Engagement (Frequency)} & \cellcolor[HTML]{FFFFC7}Global Community Density \\
6 & \cellcolor[HTML]{E5FFD6}{\color[HTML]{000000} \% Normal User Engagement (Frequency)} & \cellcolor[HTML]{ECF4FF}Median Spreader Influence (Authority) \\
7 & \cellcolor[HTML]{E5FFD6}{\color[HTML]{000000} \% Susceptible User Engagement (Frequency)} & \cellcolor[HTML]{E5FFD6}\% Normal User Engagement (\#News) \\
8 & \cellcolor[HTML]{E5FFD6}{\color[HTML]{000000} \% Normal User Engagement (\#News)} & \cellcolor[HTML]{E5FFD6}\% Susceptible User Engagement (\#News) \\
9 & \cellcolor[HTML]{E5FFD6}{\color[HTML]{000000} \% Susceptible User Engagement (\#News)} & \cellcolor[HTML]{ECF4FF}Median Spreader Influence (In-degrees) \\
10 & \cellcolor[HTML]{ECF4FF}\% Normal Spreaders (Frequency) & \cellcolor[HTML]{E5FFD6}\% Normal User Engagement (Frequency) \\
11 & \cellcolor[HTML]{ECF4FF}\% Susceptible Spreaders (Frequency) & \cellcolor[HTML]{E5FFD6}\% Susceptible User Engagement (Frequency) \\
12 & \cellcolor[HTML]{ECF4FF}\% Normal Spreaders (\#News) & \cellcolor[HTML]{ECF4FF}\% Normal Spreaders (\#News) \\
13 & \cellcolor[HTML]{ECF4FF}\% Susceptible Spreaders (\#News) & \cellcolor[HTML]{ECF4FF}\% Susceptible Spreaders (\#News) \\
14 & \cellcolor[HTML]{FFFFC7}Global Community Density & \cellcolor[HTML]{ECF4FF}Median Spreader Influence (In-closeness) \\
15 & \cellcolor[HTML]{FFFFC7}\% Egos (S $\rightarrow$ S, \#News) & \cellcolor[HTML]{FFFFC7}\% Egos (S $\rightarrow$ S, Frequency) \\
16 & \cellcolor[HTML]{FFFFC7}\% Egos (S $\rightarrow$ S, Frequency) & \cellcolor[HTML]{ECF4FF}\% Normal Spreaders (Frequency) \\
17 & \cellcolor[HTML]{E5FFD6}Avg. Normal User Engagement (\#News) & \cellcolor[HTML]{ECF4FF}\% Susceptible Spreaders (Frequency) \\
18 & \cellcolor[HTML]{FFFFC7}\% Egos (S $\rightarrow$ N, Frequency) & \cellcolor[HTML]{FFFFC7}\% Egos (S $\rightarrow$ S, \#News) \\
19 & \cellcolor[HTML]{FFFFC7}\% Egos (N $\rightarrow$ S, \#News) & \cellcolor[HTML]{ECF4FF}Median Spreader Influence (Hub) \\
20 & \cellcolor[HTML]{FFFFC7}\% Egos (N $\rightarrow$ S Frequency) & \cellcolor[HTML]{FFFFC7}{\color[HTML]{000000} \% Triads (N $\rightarrow$ S, S $\rightarrow$ S, S $\rightarrow$ N, \#News)} \\
\bottomrule[1pt]
\end{tabular}
\begin{tablenotes}
\item Blue: \textsc{More-Spreader Pattern}; Green: \textsc{Stronger-Engagement Pattern}; Yellow: \textsc{Denser-Network Pattern}
\end{tablenotes}
\end{threeparttable}
\end{table*}

\subsection{Performance Evaluation}
\label{subsec::performance}

Various supervised learners with 5-fold cross-validation were used in our experiments. The performance is evaluated using accuracy and $F_1$ score. In the following, we will first present the general performance of the proposed approach in Section \ref{subsubsec::comparativeStudy}. Based on that, the importance of patterns (see Section \ref{subsubsec::pattern}) and features (see Section \ref{subsubsec::feature}) in fake news detection is further assessed. The sensitivity of the proposed approach is evaluated to the threshold and calculation of user susceptibility in Section \ref{subsubsec::threshold}, as well as its sensitivity to how much labeled news articles are available and what proportion between two labels (true vs. fake) in Section \ref{subsubsec::dataSize}. The performance of our approach on fake news early detection is finally examined in Section \ref{subsubsec::earlyDetection}.

\subsubsection{General Performance Evaluation}
\label{subsubsec::comparativeStudy}

We experimented with various classifiers to detect fake news using our features, including Support Vector Machine (SVM), $k$-Nearest Neighbors ($k$-NN), Na\"ive Bayes (NB), Decision Trees (DT) and Random Forests (RF). The results obtained are all provided in Figure \ref{fig:classifiers}. It can be observed from the Figure \ref{fig:classifiers} that RF performs best on both datasets, achieving an accuracy and $F_1$ score of around 0.93 on PolitiFact and around 0.84 on BuzzFeed. 

Such performance is further compared with that of baselines, where the results are presented in Table \ref{tab::comparativeStudies}. 
Compared to the content-based (\cite{perez2017automatic,zhou2019content,mikolov2013efficient,le2014distributed}) and network-based models (\cite{castillo2011information}) among baselines, the proposed approach can perform relatively well on both datasets.
Compared to the hybrid one (\cite{shu2019beyond}), the proposed approach can be comparable with it and can outperform it when introducing the linguistic features in the proposed approach (``Our Approach + \cite{zhou2019content}'').

\subsubsection{Performance of Fake News Patterns} \label{subsubsec::pattern}
We further analyze the performance of each fake news pattern and their combinations on fake news detection. The results are presented in Table \ref{tab::patternPerformance}, which supports the following observations.
First, \textsc{More-Spreader Pattern} and \textsc{Stronger-Engagement Pattern} perform best compared to the others when being separately utilized to detect fake news, achieving around 89\% (81\%) accuracy and $F_1$ score using PolitiFact (BuzzFeed) data. The performance of \textsc{Denser-Network Pattern} follows.
Second, when combining different patterns, their performance is in general better than when separately using them, which can achieve an accuracy and $F_1$ score of around 93\% (82\%) on PolitiFact (BuzzFeed).
Third, when using all patterns to detect fake news, a significantly better performance is achieved compared to using network similarity, which provides a mix of patterns from a higher network view, which is a positive sign for our summarized fake news patterns and defined representations of patterns in networks.
Fourth, network similarity features can slightly improve the performance of the combination of four fake news propagation patterns, which finally achieves an accuracy and $F_1$ score of around 93\% (84\%) on PolitiFact (BuzzFeed).

\footnotetext{The results are based on Weisfeiler-Lehman graph kernel~\cite{shervashidze2011weisfeiler}. As a widely-accept graph kernel, Weisfeiler-Lehman graph kernel can measure the similarities among labeled graphs, which we treat TNNs and FNNs as.}

\subsubsection{Feature Importance Analysis}
\label{subsubsec::feature}

\begin{figure}[h]
    \centering
    \subfigure[PolitiFact]{
    \begin{minipage}{0.22\textwidth}
    \includegraphics[width=\textwidth]{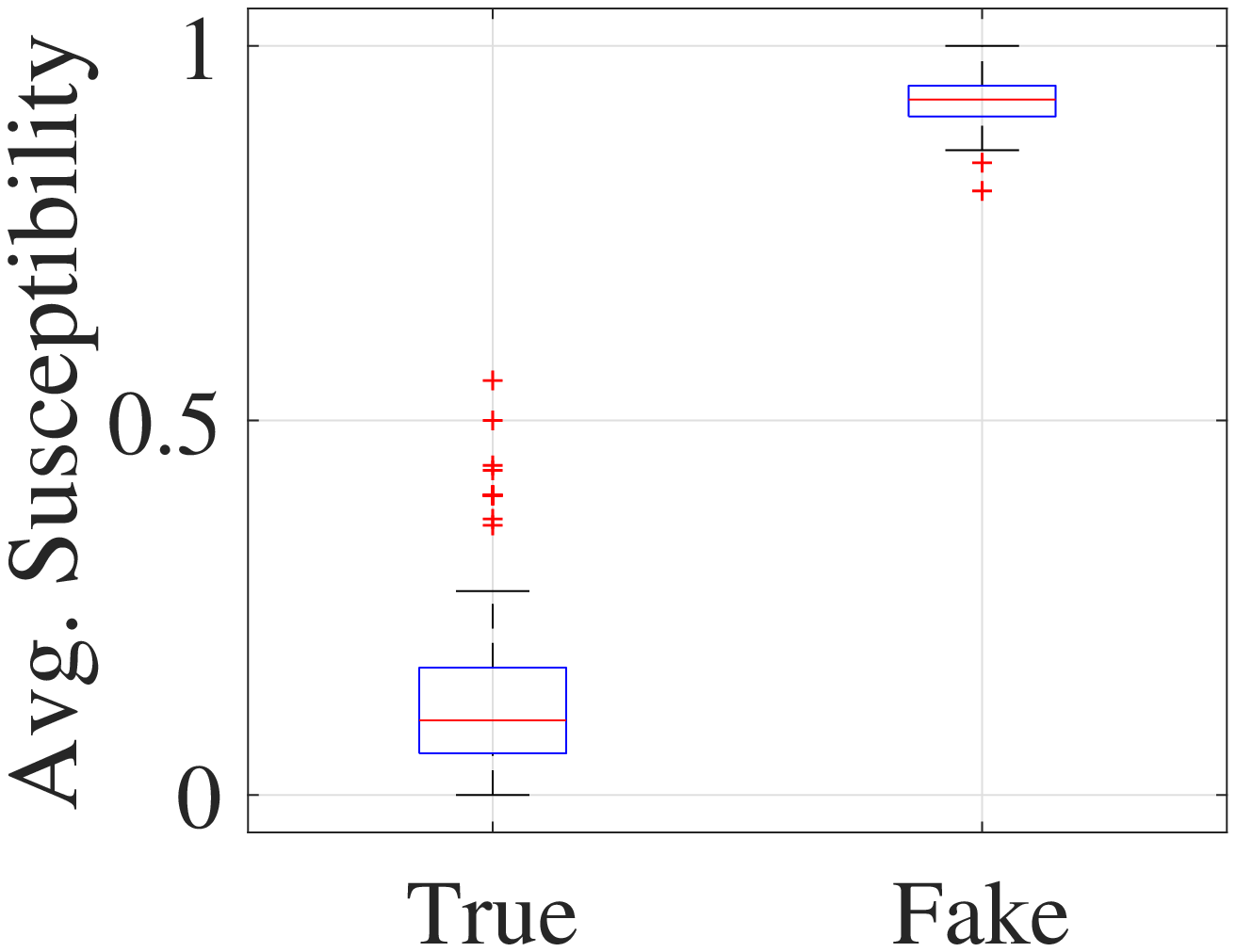}
    \includegraphics[width=\textwidth]{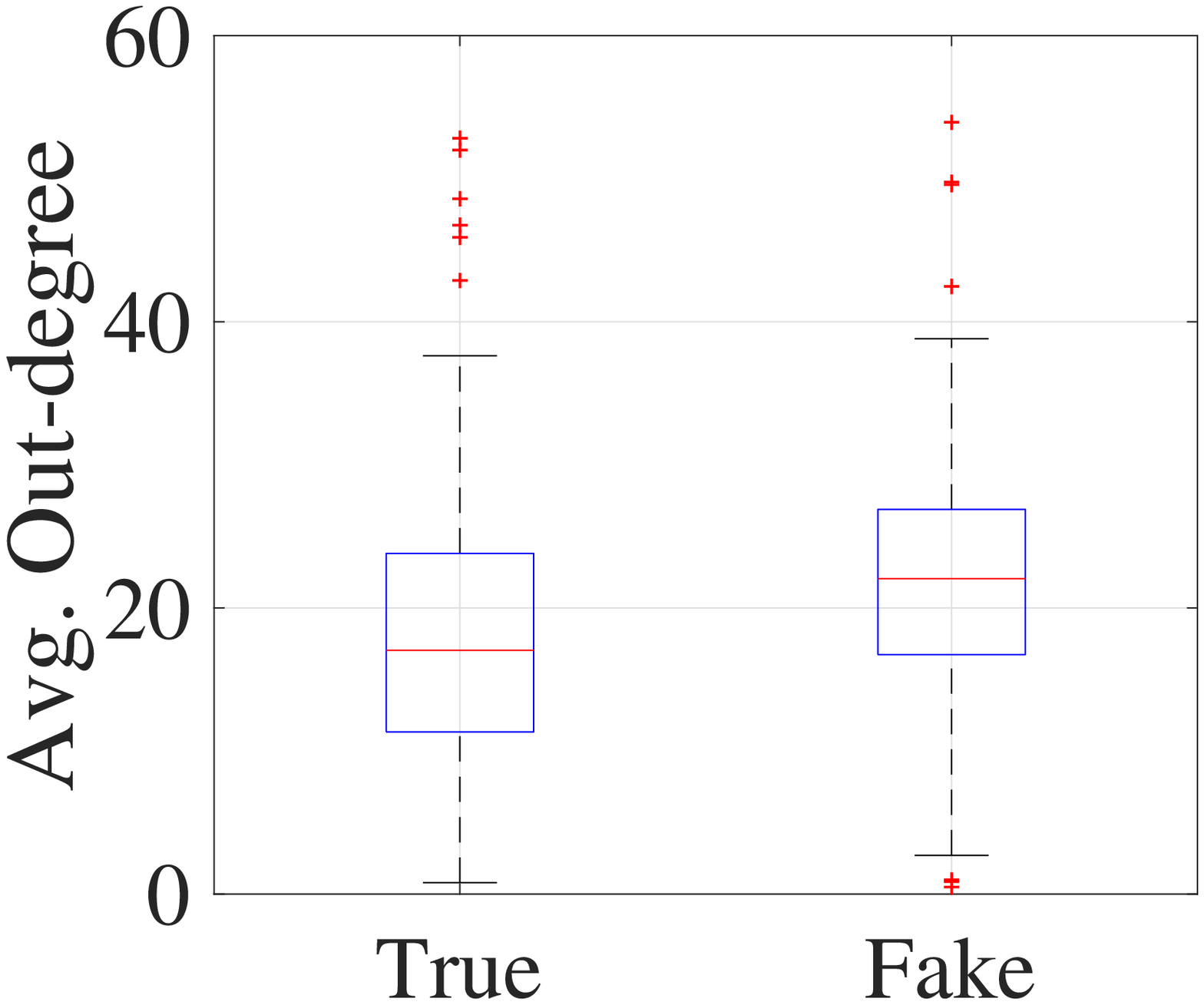}
    \includegraphics[width=\textwidth]{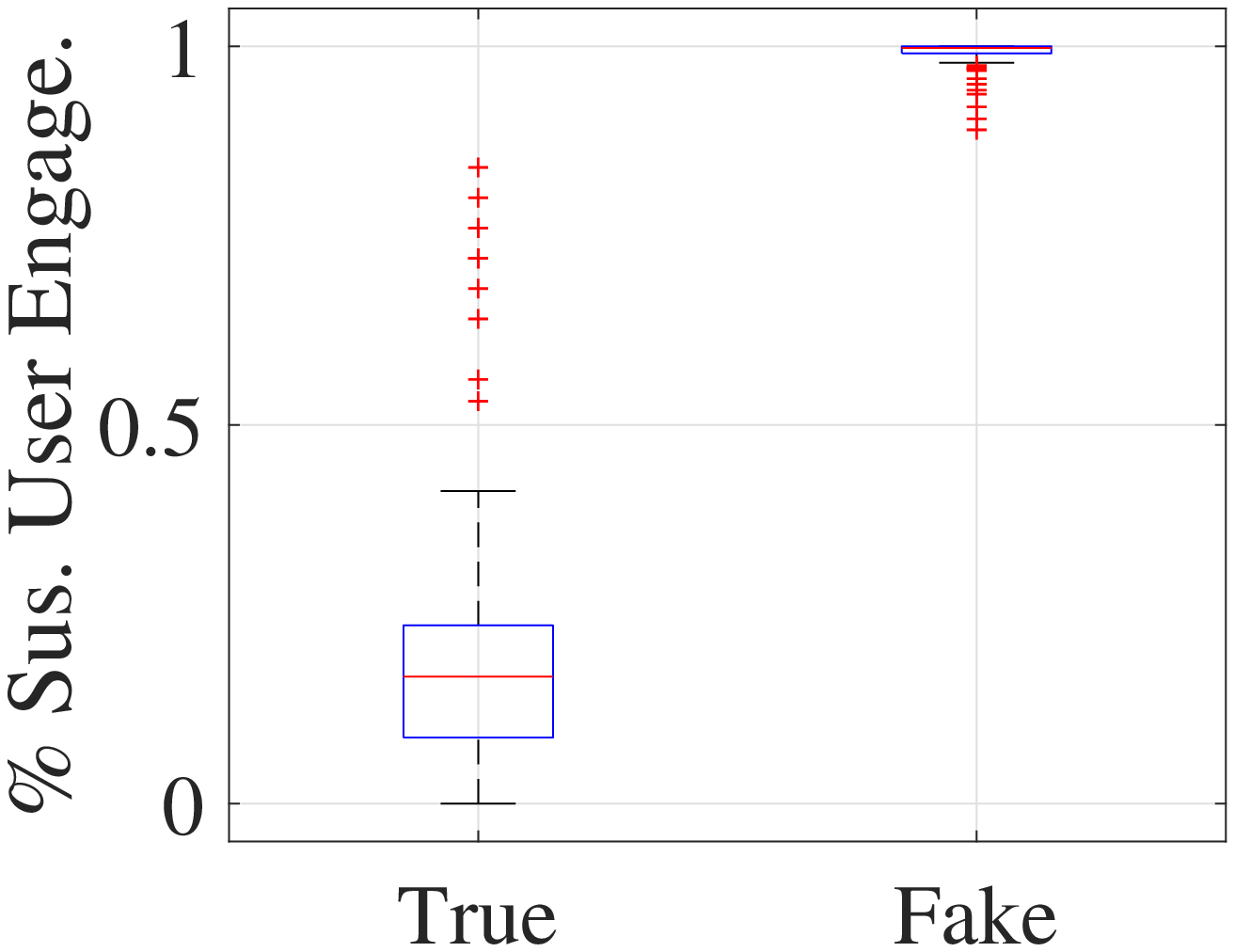}
    \includegraphics[width=\textwidth]{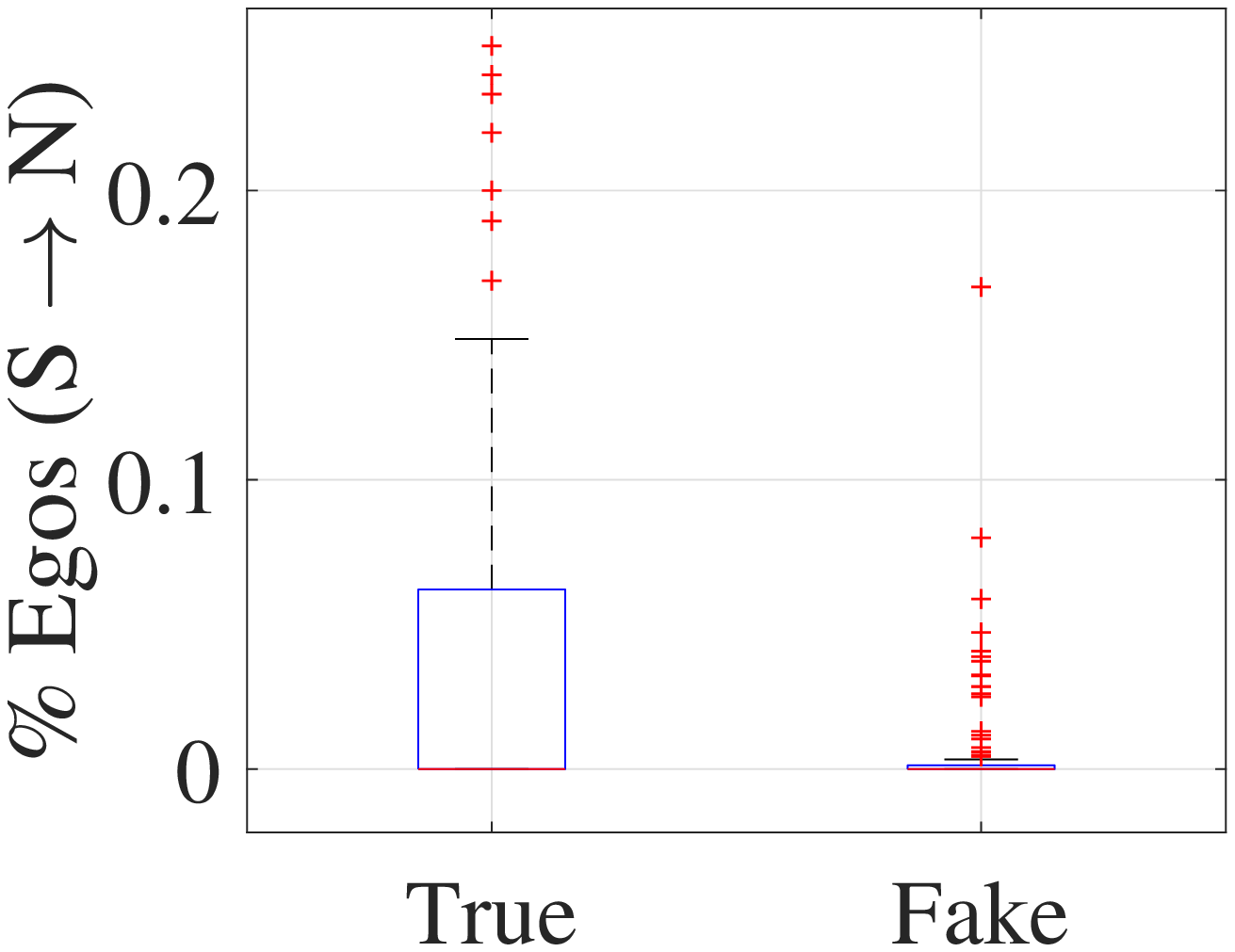}

    \includegraphics[width=\textwidth]{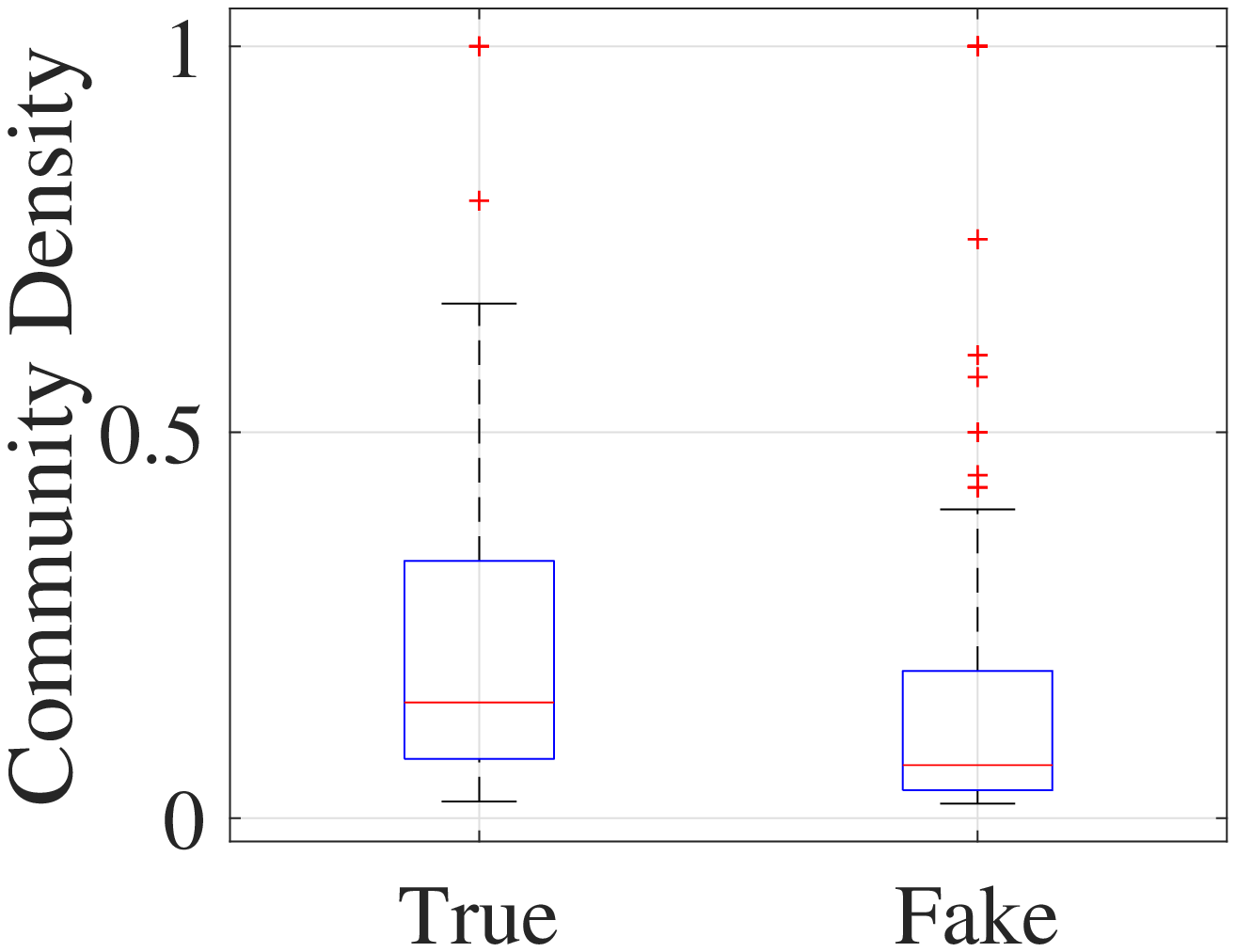}
    \end{minipage}}
    \subfigure[BuzzFeed]{
    \begin{minipage}{0.22\textwidth}
    \includegraphics[width=\textwidth]{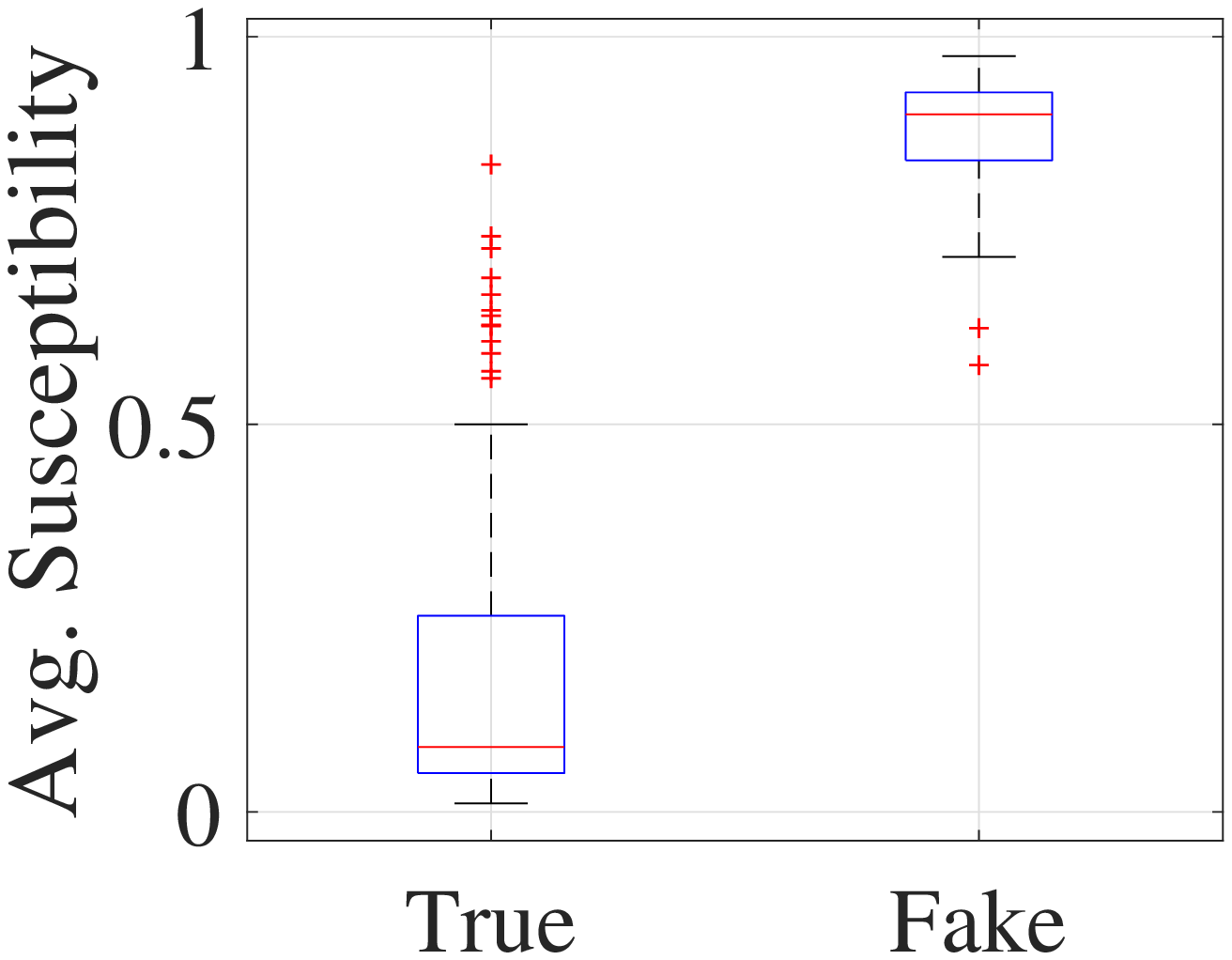}
    \includegraphics[width=\textwidth]{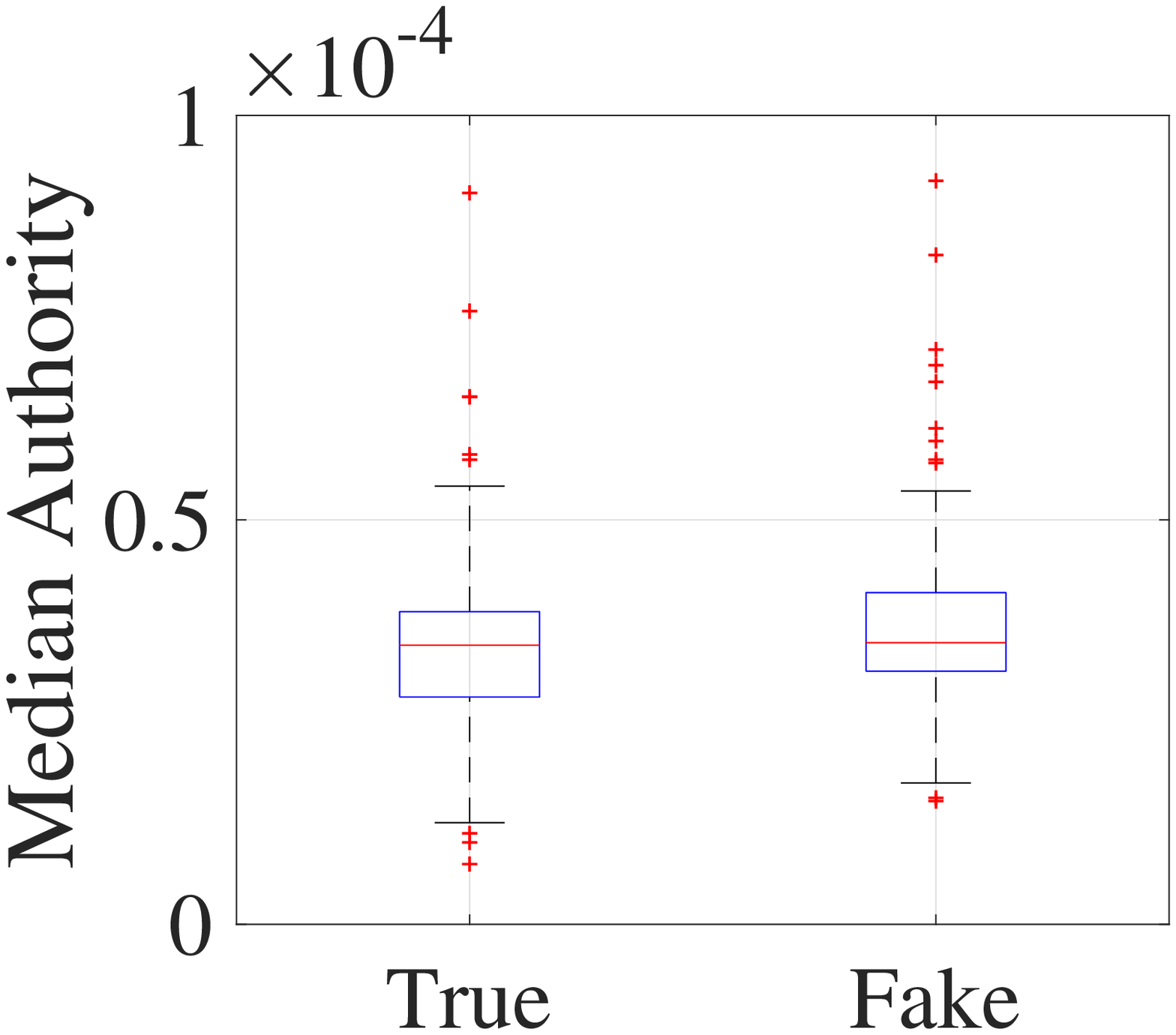}
    \includegraphics[width=\textwidth]{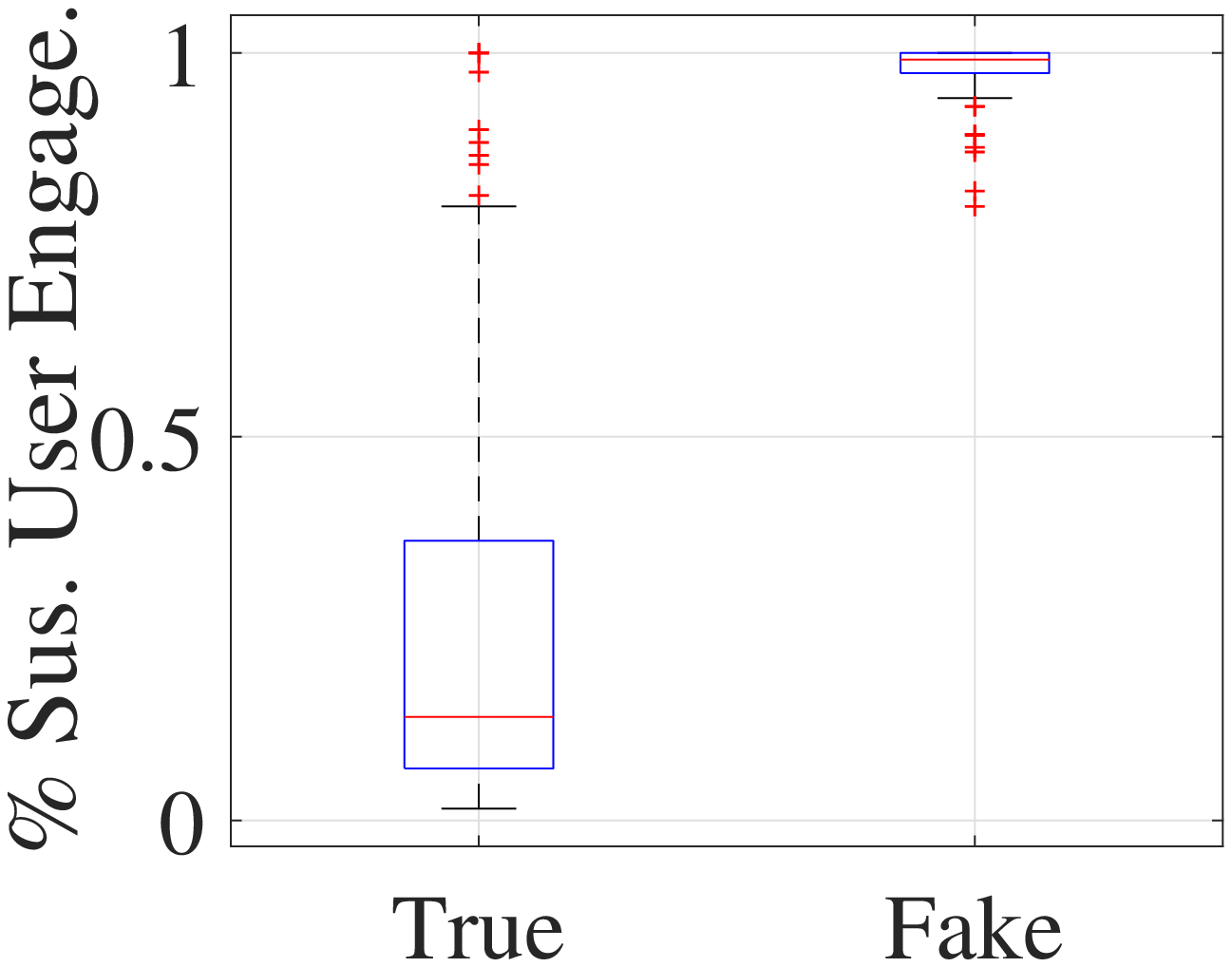}
    \includegraphics[width=\textwidth]{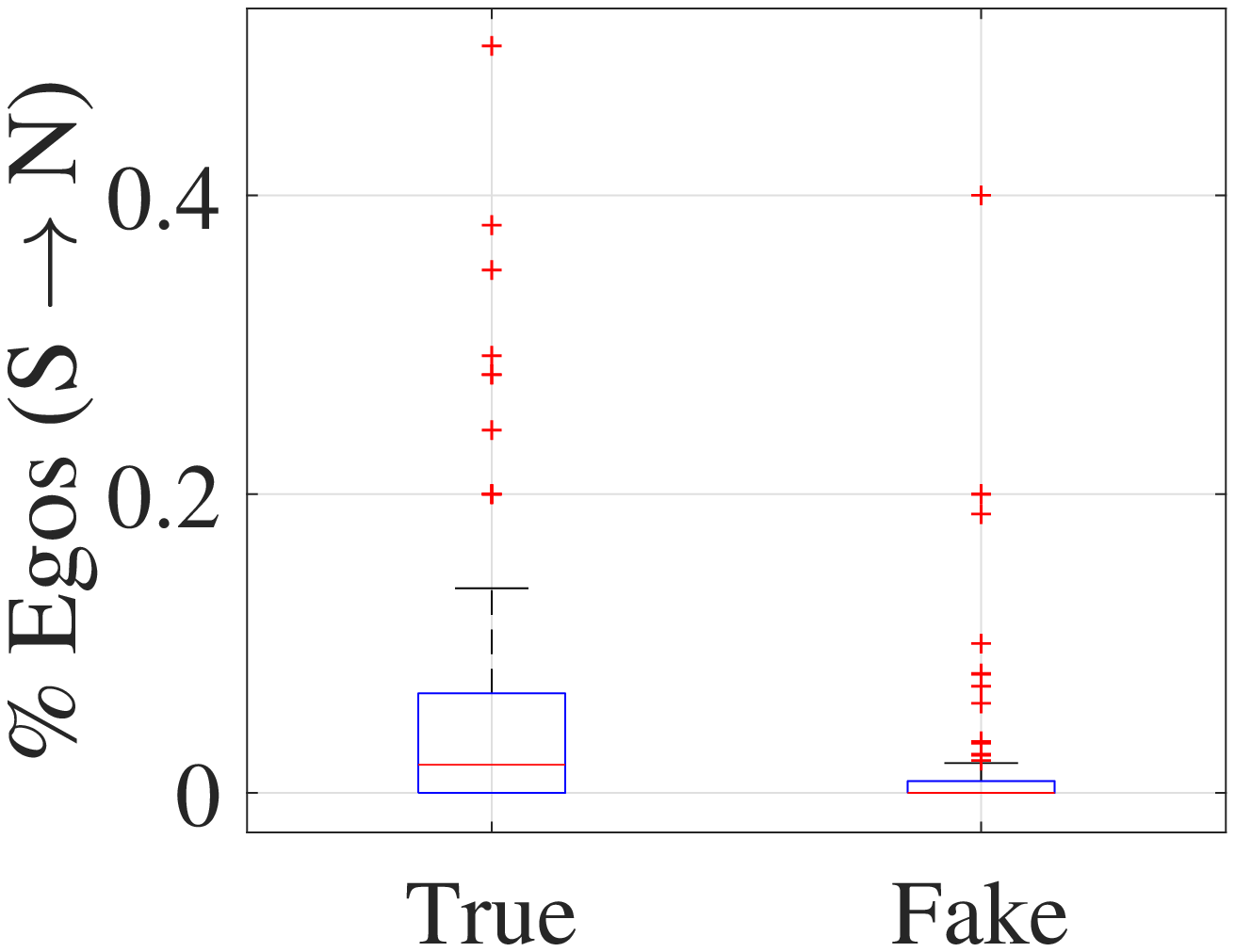}
    \includegraphics[width=\textwidth]{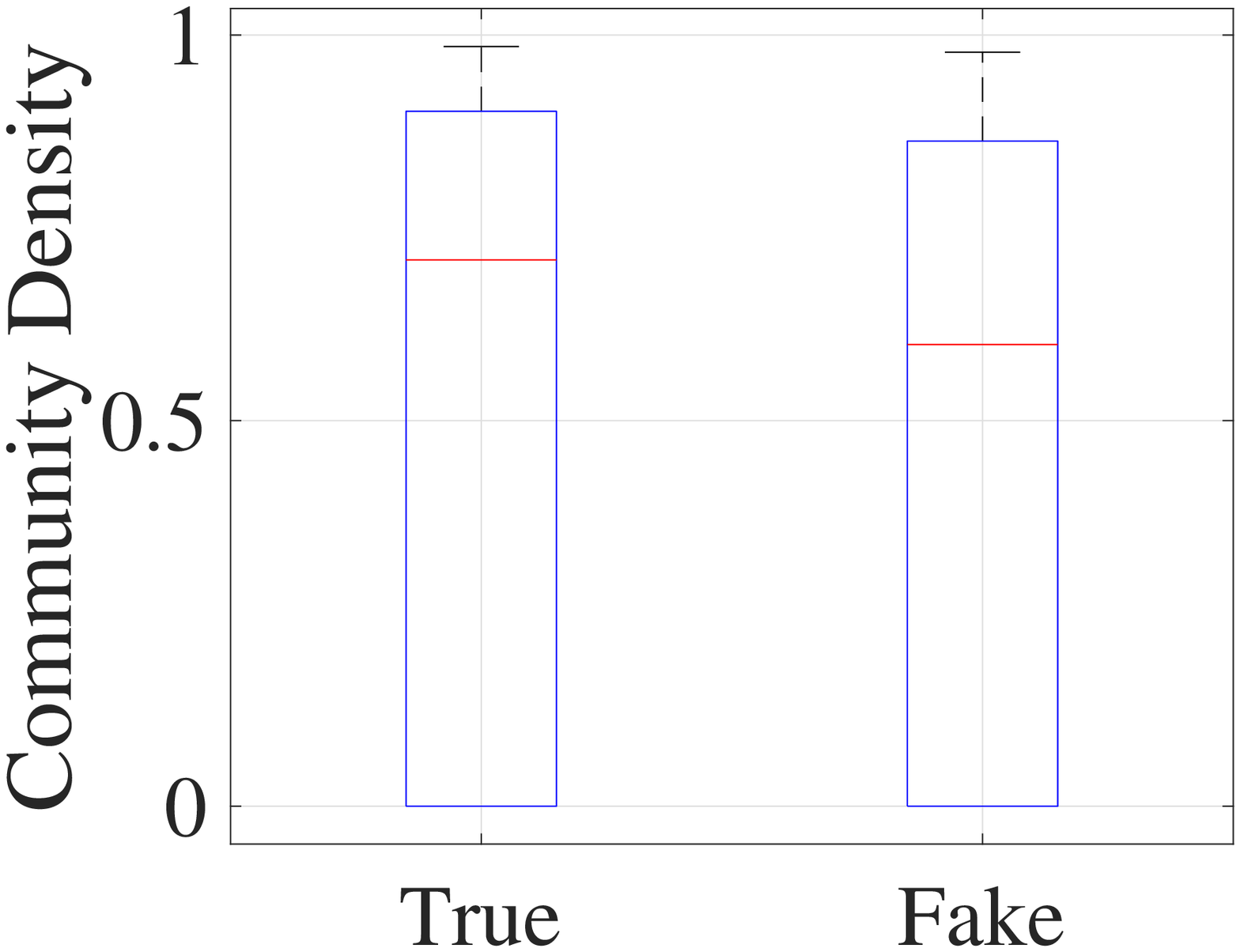}
    \end{minipage}}
    \caption{Statistics of Illustrated Important Features for Fake and True News
    }
    \label{fig:discriminatedFeatures}
\end{figure}

Features are ranked by their importance in fake news detection. Results are shown in Table~\ref{tab::top20}, which are obtained by Relief algorithm, a widely-accept feature selection algorithm~\cite{kira1992practical}.
Consistent with the performance of patterns, features representing \textsc{More-Spreader Pattern}, \textsc{Stronger-Engagement Pattern}, and \textsc{Denser-Network Pattern} are relatively more discriminative in predicting fake news compared to the other features. In addition, it can be observed from Table~\ref{tab::top20} and Figure~\ref{fig:discriminatedFeatures} that for \textsc{More-Spreader Pattern}, features contributing most to fake news detection are the mean or median of (i) spreader susceptibility and (ii) spreader influence, where fake news spreaders often share a higher susceptibility and centrality score compared to true news spreaders. For \textsc{Stronger-Engagement Pattern}, such features relate to (iii) susceptible and normal user engagements, where susceptible (normal) users engage more strongly in fake (true) news compared to true (fake) news. For \textsc{Denser-Network Pattern}, such features are generally at (iv) ego and (v) community level. Specifically, FNNs are characterized with a higher proportion of connections between susceptible users ($S \rightarrow S$) while TNNs are characterized with a higher proportion of other ego relations ($S \rightarrow N$, $N \rightarrow S$ and $N \rightarrow N$). With a same network size, a FNN often has less communities compared to a TNN, which indicate that a denser network structure is often within a FNN compared to a TNN.

\begin{figure}[t]
    \centering
    \includegraphics[width=0.48\textwidth]{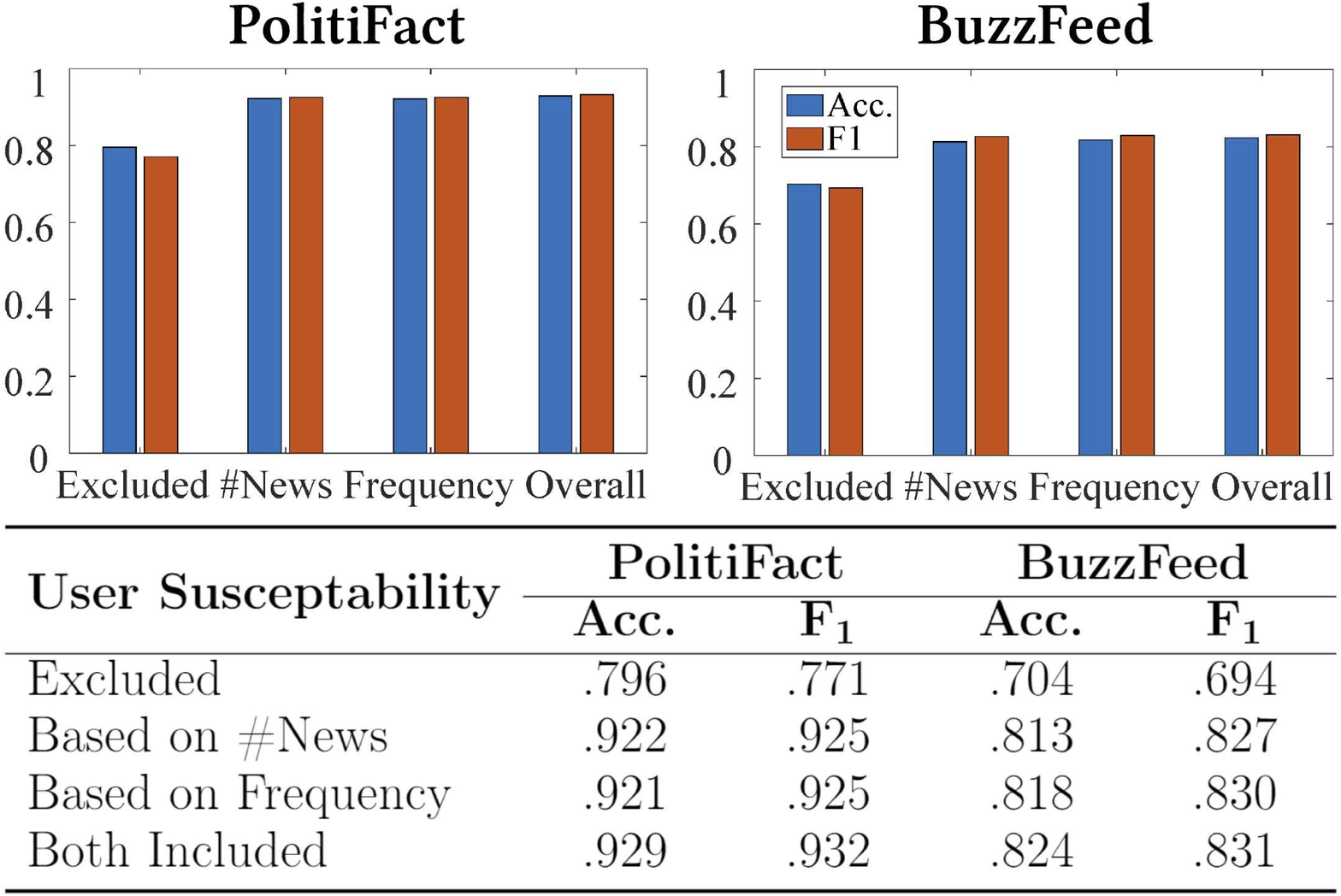}
    \caption{Impact of the Utilizing and Means to Calculate User Susceptibility on Fake News Detection.  Considering user susceptibility can improve fake news prediction, while two methods of computing user susceptibility perform similarly.}
    \label{fig:userSusceptibility}
\end{figure}

\begin{figure}
    \centering
    \subfigure[PolitiFact]{
    \includegraphics[width = 0.47\textwidth]{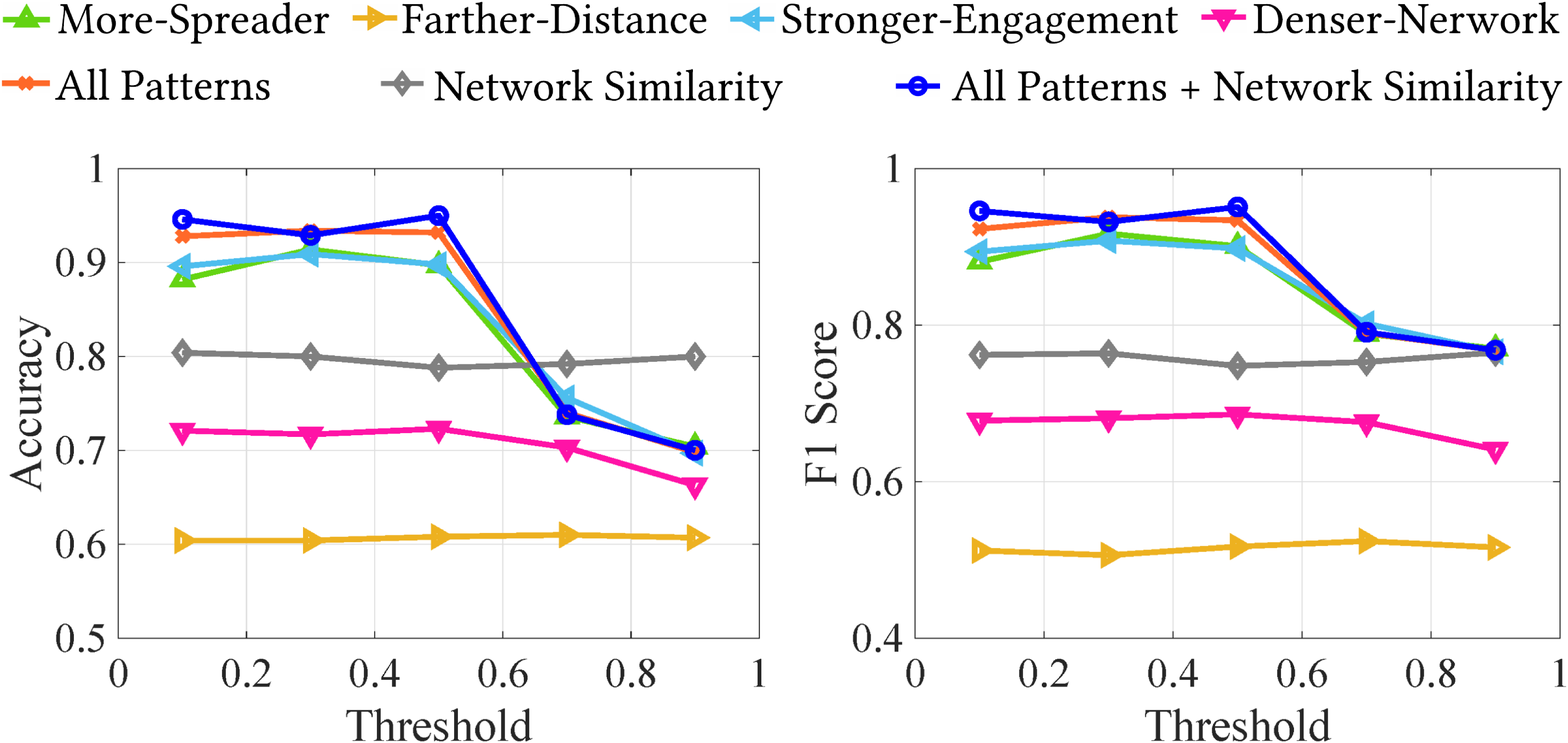}
    }
    \subfigure[BuzzFeed]{
    \includegraphics[width = 0.47\textwidth]{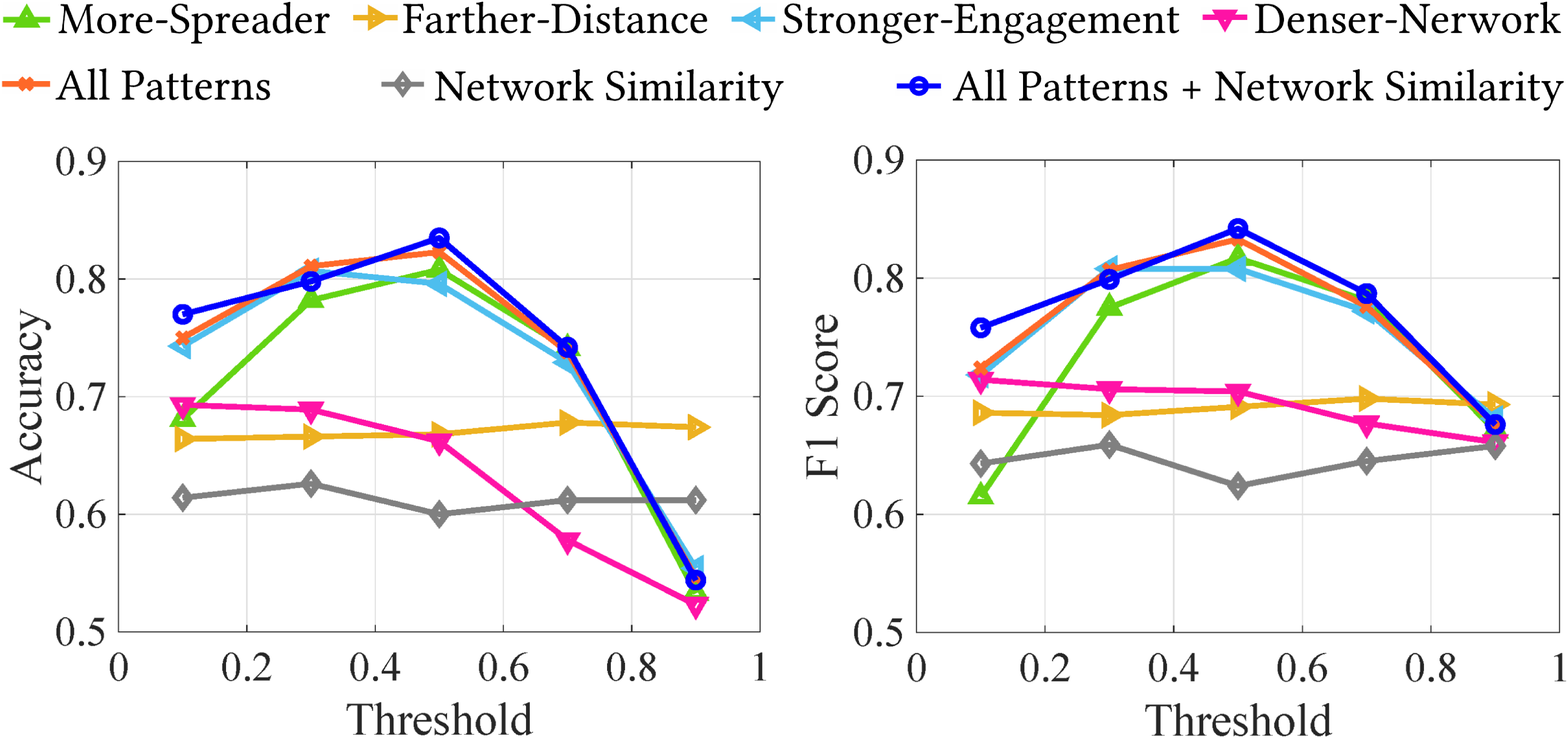}
    }
    \caption{Impact of User Susceptibility Threshold on Fake News Detection. When the threshold changes, (i) the performance of \textsc{Farther-Distance Pattern} and network similarity is hardly impacted due to no relevance to user susceptibility; (ii) \textsc{More-Spreader Pattern} or \textsc{Stronger-Engagement Pattern} can outperform \textsc{Denser-Network Pattern} though they are less stable; (iii) using all patterns (with/without network similarity) can always perform comparatively well compared to the others and achieve the highest performance when the threshold value is 0.5.}
  \label{fig::threshold}
\end{figure}

\begin{figure*}[t]
    \centering
    \subfigure[PolitiFact]{
    \begin{minipage}{0.48\textwidth}
    \includegraphics[width = 0.5\textwidth]{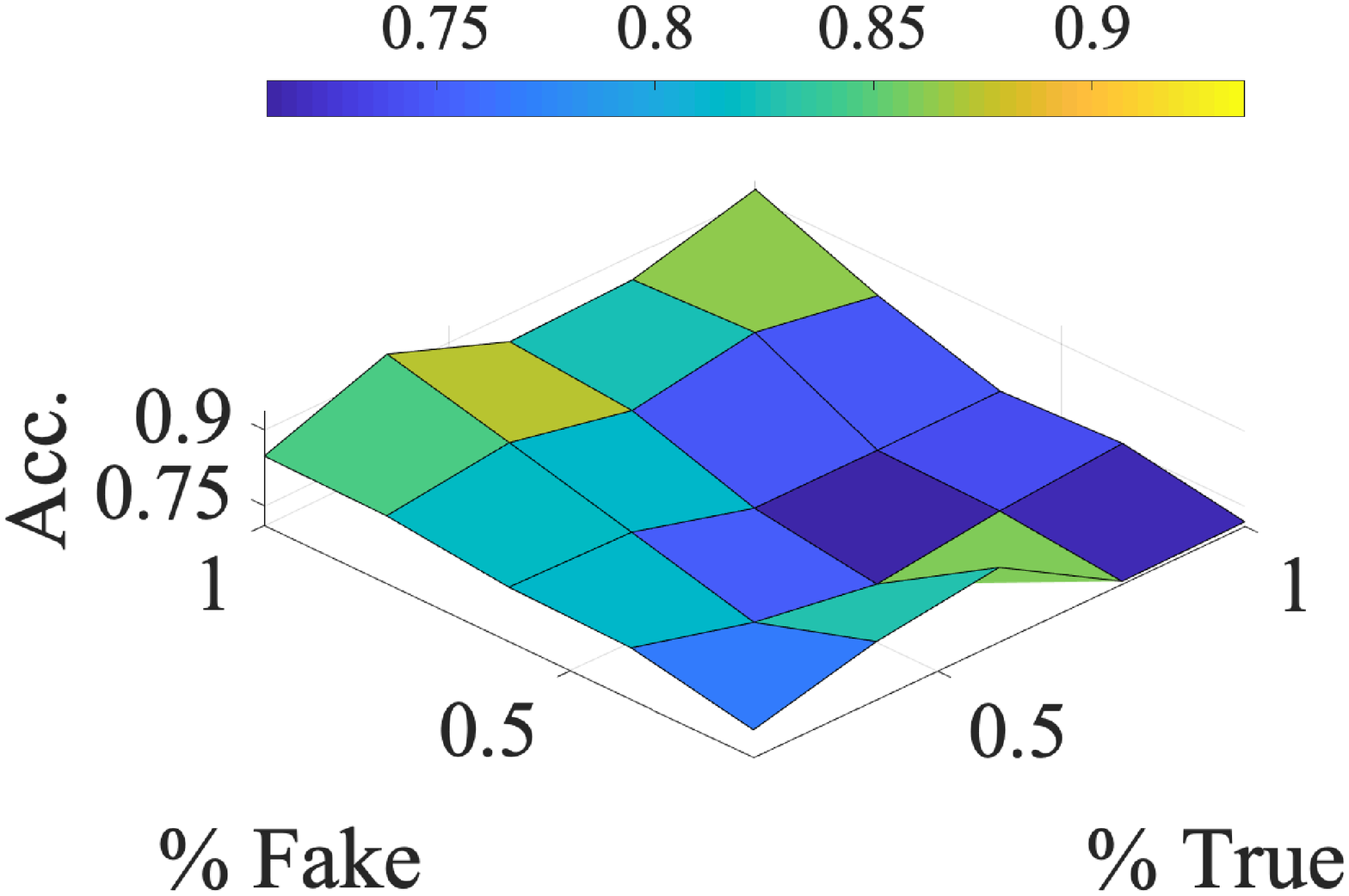}
    \includegraphics[width = 0.5\textwidth]{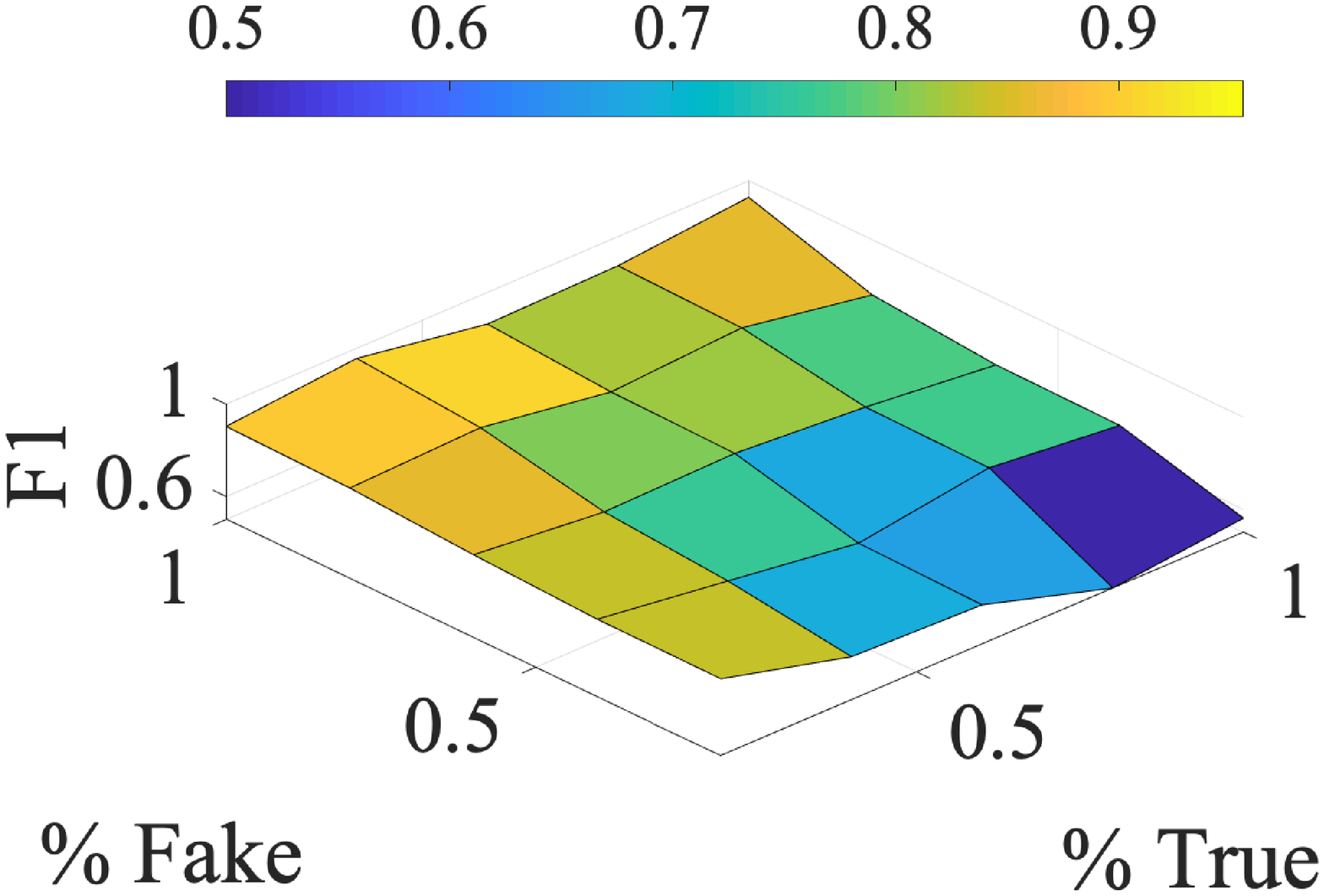}
    \includegraphics[width = \textwidth]{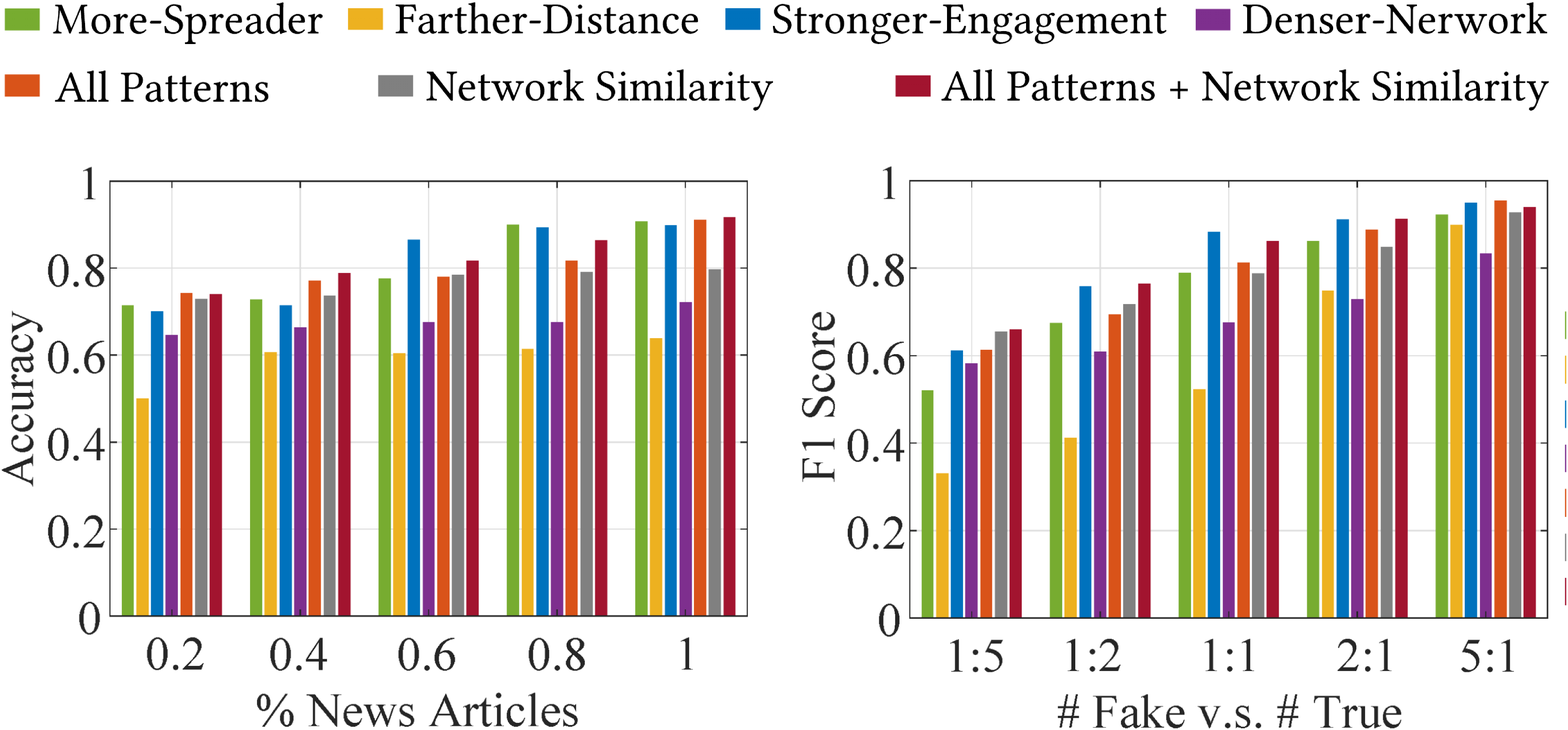}
    \end{minipage}
    }
    \subfigure[BuzzFeed]{
    \begin{minipage}{0.48\textwidth}
    \includegraphics[width = 0.5\textwidth]{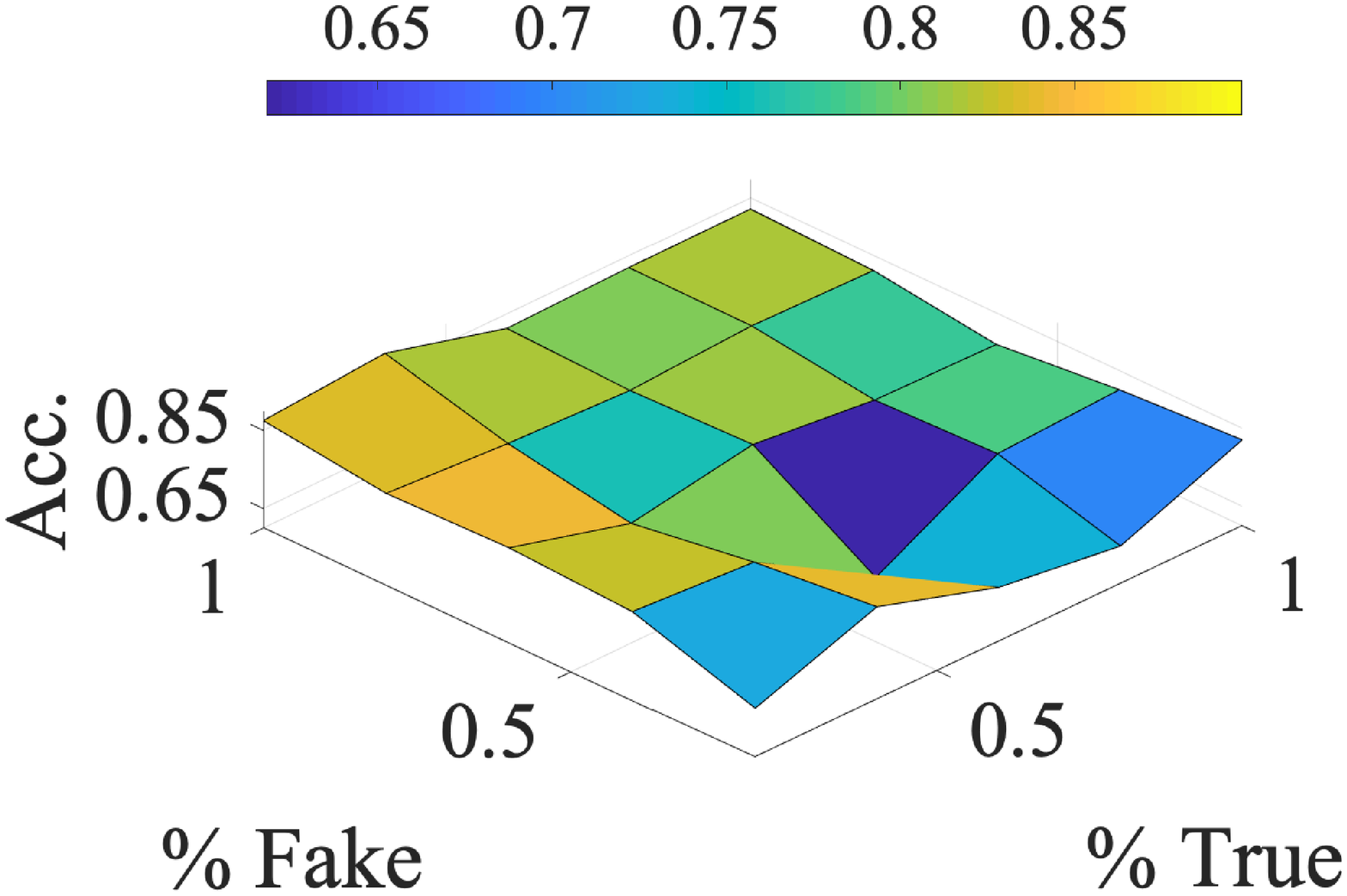}
    \includegraphics[width = 0.5\textwidth]{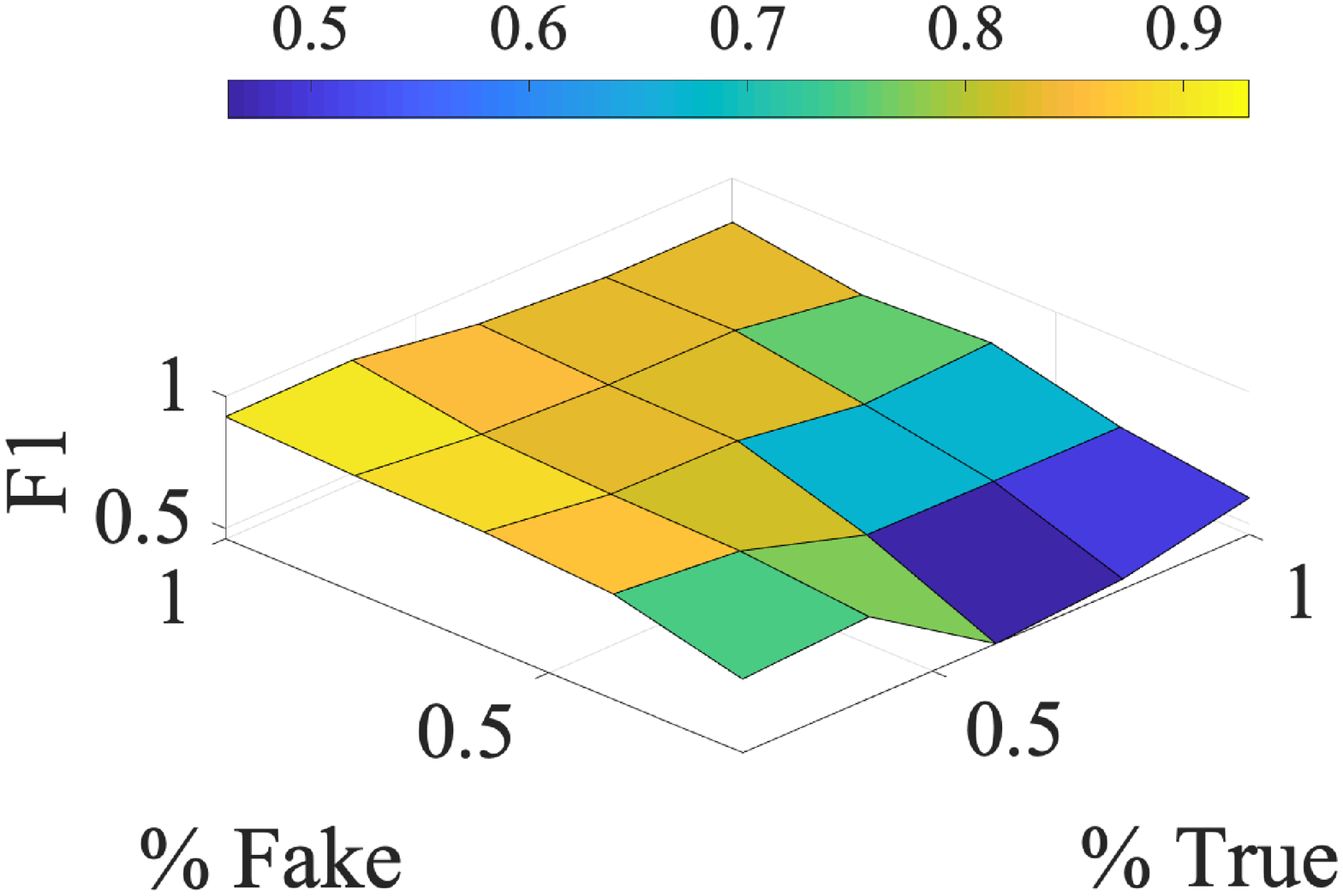}
    \includegraphics[width = \textwidth]{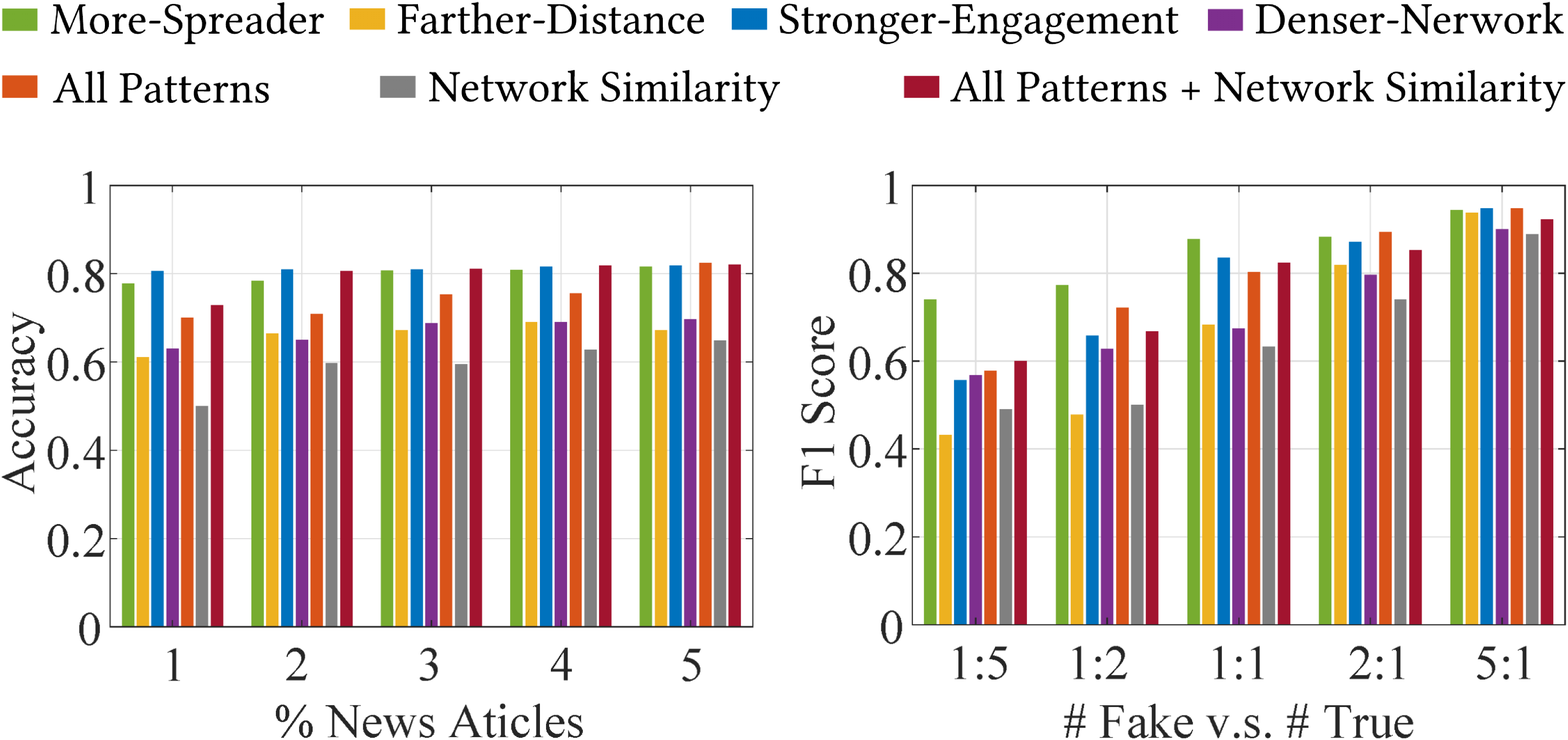}
    \end{minipage}
    }
    \vspace{-4mm}
    \caption{Impact of News Number and Distribution on Fake News Prediction. In general, (i) the proposed approach can achieve an accuracy rate (a $F_1$ score) $\sim$0.7 ($\sim$0.65) to $\sim$0.85 ($\sim$0.9) in most cases on both datasets (see the upper four figures). When only the number of news articles varies, an accuracy rate (here the datasets are class-balanced) between $\sim$0.73 ($\sim$0.8) to $\sim$0.9 ($\sim$0.82) can be achieved on PolitiFact (BuzzFeed) data and overall features (\textsc{Stronger-Engagement Pattern}). When the news distribution varies, a $F_1$ score (here the dataset can be unbalanced) ranging from $\sim$0.65 ($\sim$0.75) to $\sim$0.93 ($\sim$0.92) can be achieved on PolitiFact (BuzzFeed) data and overall features (\textsc{More-Spreader Pattern}).}
  \label{fig:newsDistribution}
\end{figure*}

 \begin{figure}[t]
    \centering
    \subfigure[PolitiFact]{
    \includegraphics[width = 0.235\textwidth]{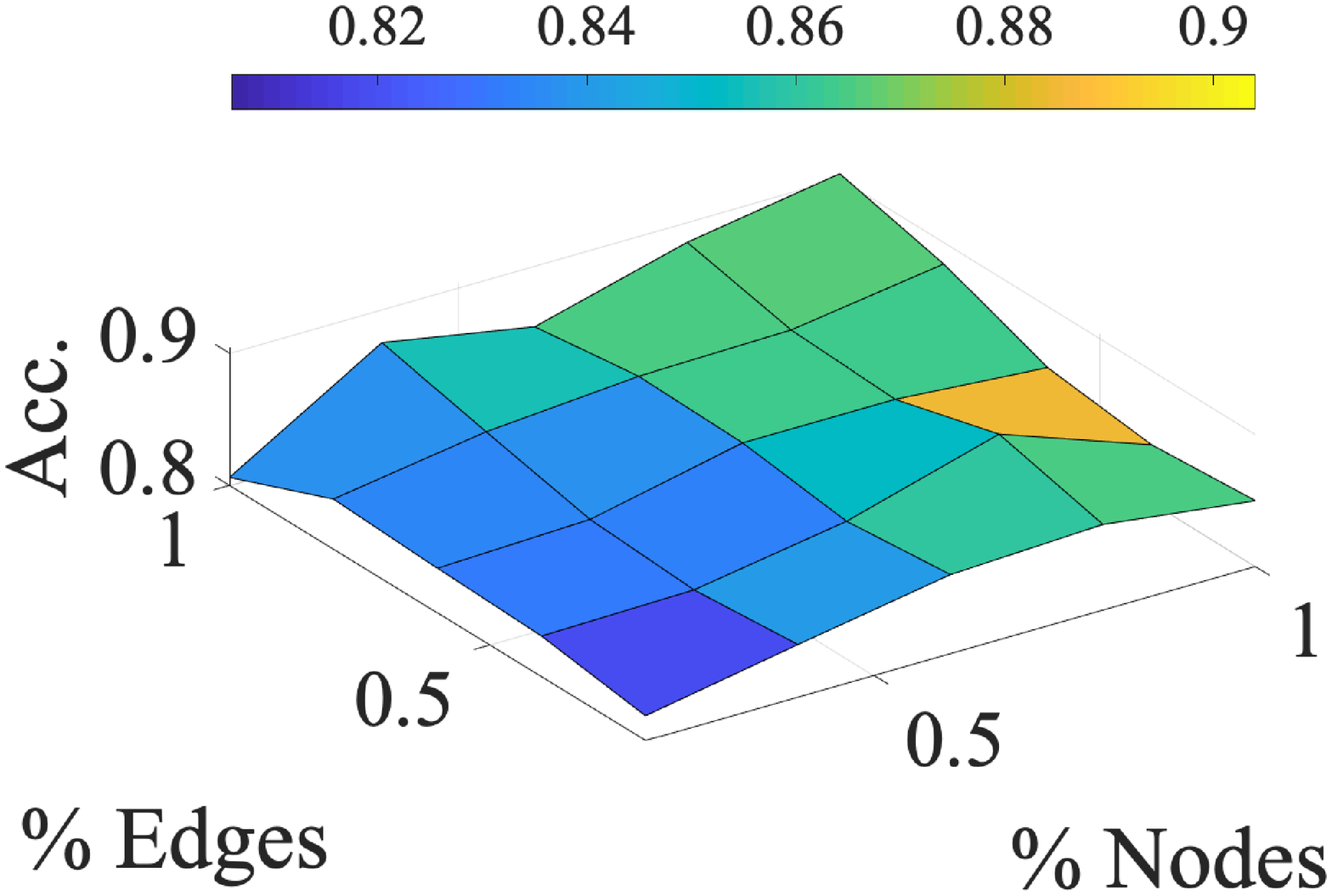}
    \includegraphics[width = 0.235\textwidth]{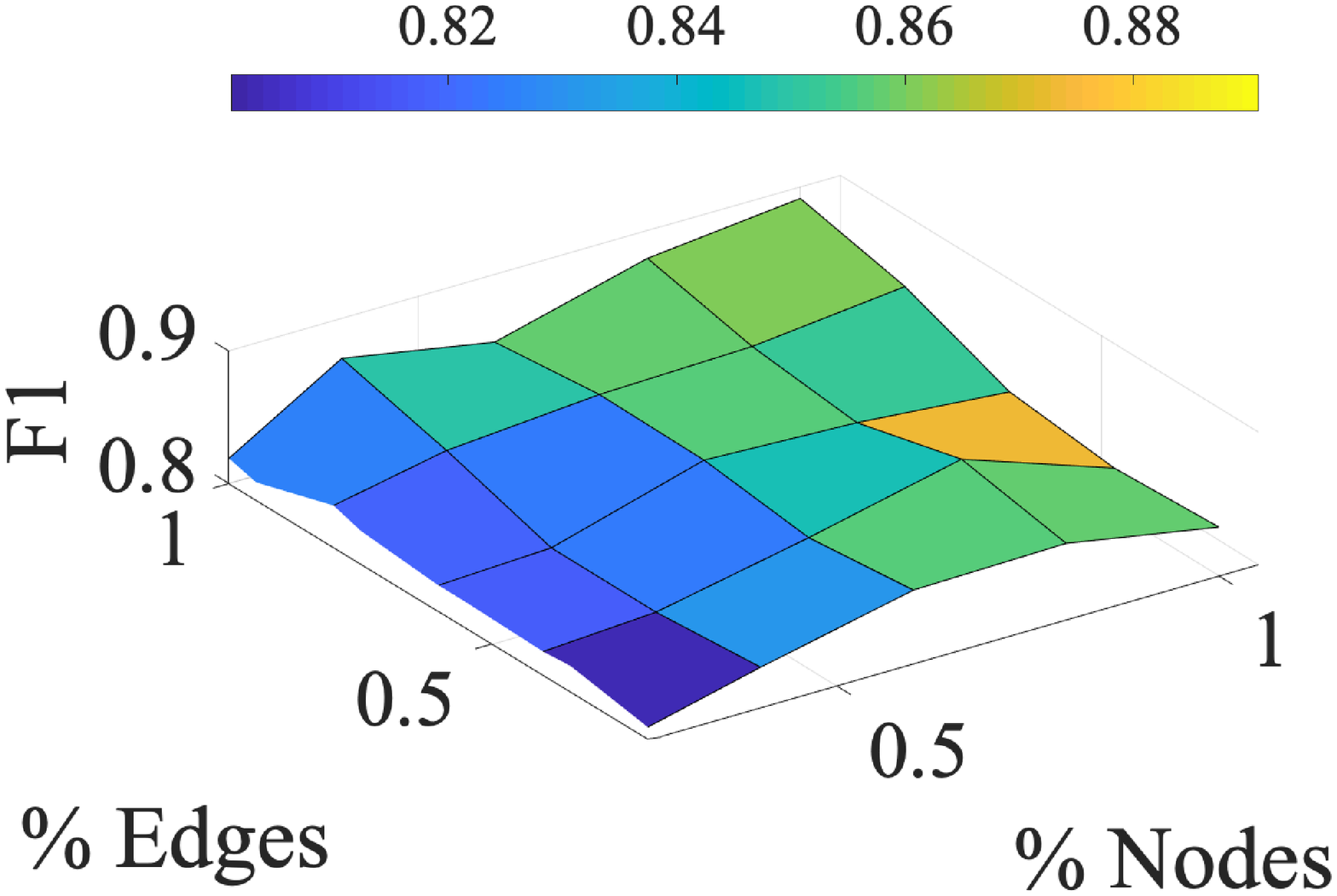}}
    \subfigure[BuzzFeed]{
    \includegraphics[width = 0.235\textwidth]{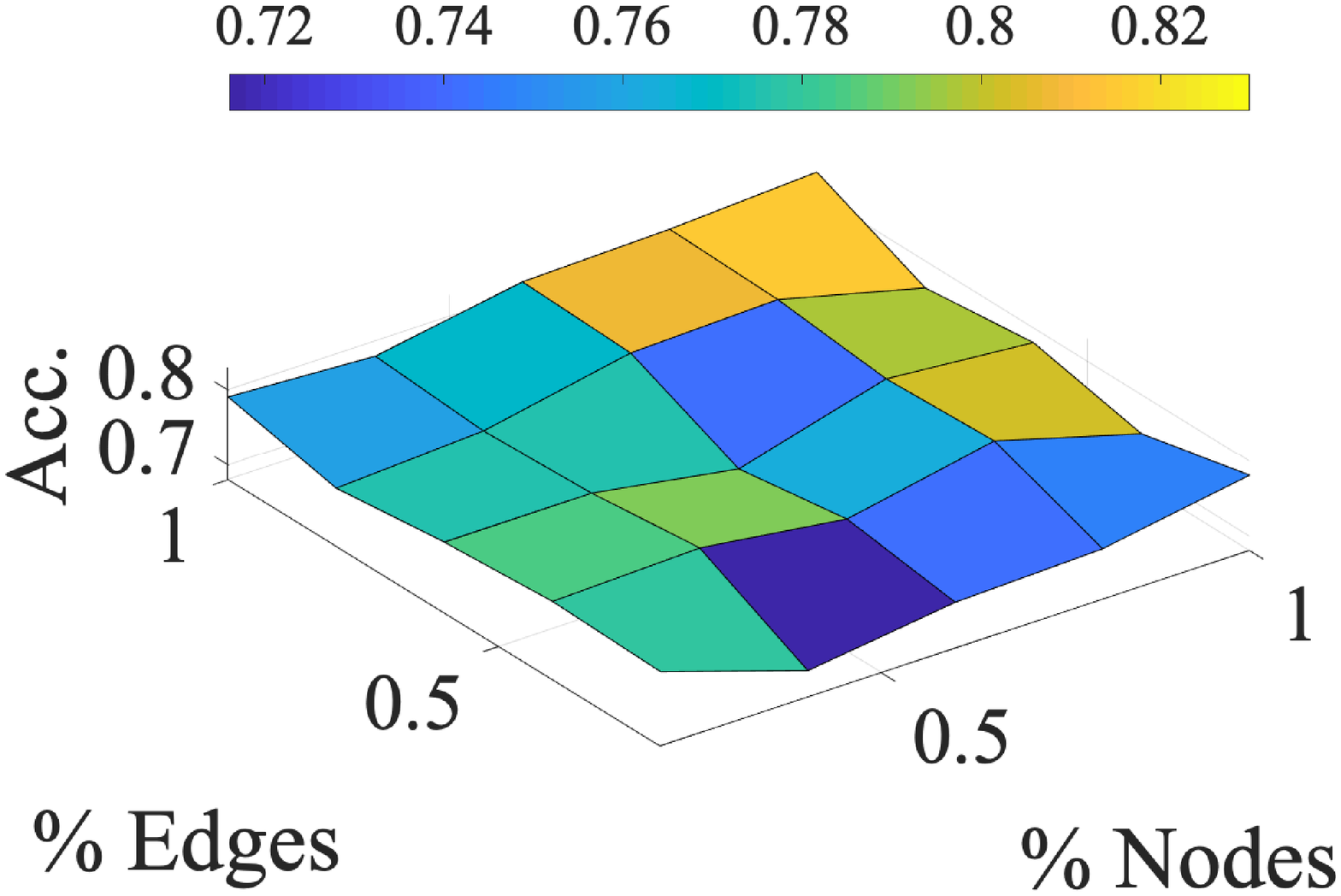}
    \includegraphics[width = 0.235\textwidth]{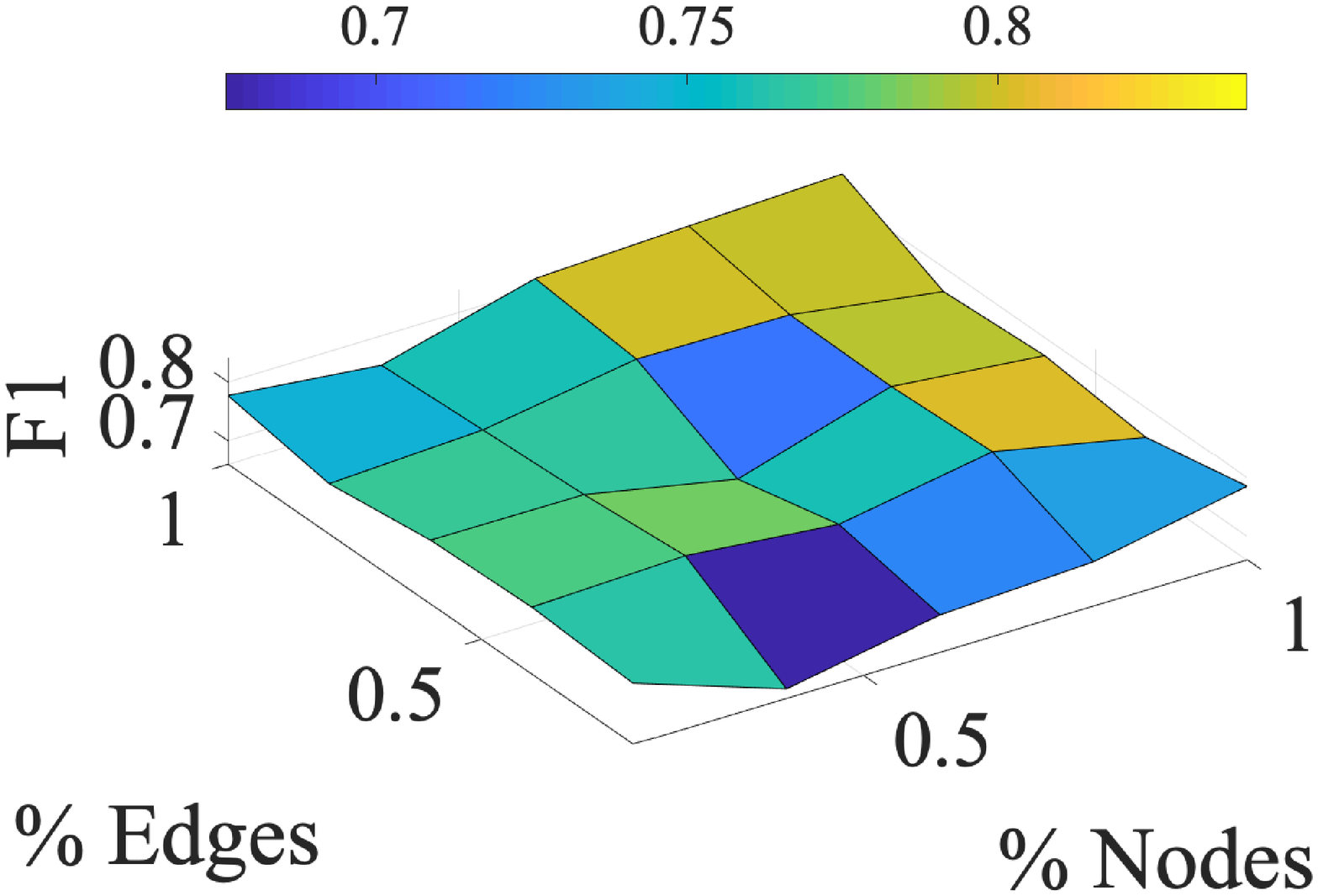}}
    \vspace{-4mm}
    \caption{Impact of Available Network Information on Fake News Prediction. The proposed approach is generally stable with an accuracy and $F_1$ score between $\sim$0.8 ($\sim$0.7) and $\sim$0.9 ($\sim$0.82) on PolitiFact (BuzzFeed) data.}
  \label{fig::earlyDetection}
\end{figure}

\subsubsection{User Susceptibility Analysis}
\label{subsubsec::threshold}
Two methods have been defined to compute user susceptibility [to fake news] which plays an important role in representing patterns - one is based on the number of fake news that a user has spread (Equation (\ref{eq::userVirality1})) and the other is based on the frequency of a user on spreading fake news (Equation (\ref{eq::userVirality2})).
Once such susceptibility is computed, whether a user is susceptible or normal relies on the selection of a threshold. Thus, here we assess the impact of user susceptibility on fake news prediction by (I) how the susceptibility is computed and (II) how the threshold is determined.

\vspace{0.5em}
\noindent \textbf{I. Computation of User Susceptibility.} 
To evaluate the impact of the [non-] existence of user susceptibility and how it is computed, we conduct fake news detection (i) without user susceptibility features, and by using features that compute user susceptibility based on the (ii) number of news spread, (iii) frequency of the spreading, and (iv) both number and frequency. The results are provided in Figure \ref{fig:userSusceptibility}. It can be observed that utilizing user susceptibility can enhance the performance by $\sim$10\% when predicting fake news based on both PolitiFact and BuzzFeed datasets. However, no significant performance difference exists between the two ways that user susceptibility can be calculated.

\vspace{0.5em}
\noindent \textbf{II. Evaluating User Susceptibility Threshold.} To evaluate the impact of susceptibility threshold on fake news prediction, we set the threshold value from 0 (i.e., all users are susceptible) to 1 (i.e., all users are normal) and use the proposed approach to predict fake news based on different threshold values. The results are plotted in Figure \ref{fig::threshold}. It can be observed that when the threshold changes, (i) as the features representing \textsc{Farther-Distance Pattern} and network similarity do not need to compute susceptibility scores of users, the performance is rarely impacted; (ii) \textsc{More-Spreader Pattern} or \textsc{Stronger-Engagement Pattern} can outperform \textsc{Denser-Network Pattern} while are less stable; (iii) the combination of all patterns (with/without network similarity) can always perform comparatively well compared to the others and achieves the highest performance when the threshold value is 0.5.

 
\subsubsection{Impact of News Number and Distribution}
\label{subsubsec::dataSize}
As in the practice, the number and distribution (the proportion between fake and true news) of news articles on social networks can be dynamic and change, here we evaluate the impact of the (i) number and (ii) distribution of news articles available on the performance of the proposed method. To that end, a certain proportion ($\in [0,1]$) of samples is randomly selected from the population of true news stories in a dataset and that of fake news stories in that dataset, respectively. The performance of the proposed approach with each proportion of true and fake news is plotted in Figure \ref{fig:newsDistribution} (the upper row). Results in Figure \ref{fig:newsDistribution} indicate that, in general, the proposed approach can perform an accuracy rate of $\sim$0.7 to $\sim$0.85 and an $F_1$ score of $\sim$0.65 to $\sim$0.9 in most cases on both datasets.

Note that two variables (i.e., the number and distribution of news articles) both exist and change in this process. For a clear observation, we first control the sampled news distribution to be the same as that in original datasets, and record the performance of the proposed method with various number of overall news articles available for training and predicting fake news. On the other hand, we keep the a fixed number of news articles while vary the proportion between fake and true news in it.

Results are all provided in Figure \ref{fig:newsDistribution} (the lower row). It can be observed that (i) the impact of the number of news articles is less significant compared to the news distribution when predicting fake news based on the proposed method; (ii) when varying the number of news articles, an accuracy rate (we only present the accuracy performance in Figure \ref{fig:newsDistribution} as the datasets are balanced at this time) between $\sim$0.73 ($\sim$0.8) to $\sim$0.9 ($\sim$0.82) can be achieved by using PolitiFact (BuzzFeed) data and all patterns plus network similarity (\textsc{Stronger-Engagement Pattern}); and (iii) when varying the news distribution, a $F_1$ score (we evaluate the performance only based on $F_1$ score here as the datasets can be unbalanced) ranging from $\sim$0.65 ($\sim$0.75) to $\sim$0.93 ($\sim$0.92) can be achieved by using PolitiFact (BuzzFeed) data and all patterns plus network similarity (\textsc{More-Spreader Pattern}).

\subsubsection{Early Detection Analysis}
\label{subsubsec::earlyDetection}

Fake news \textit{early} detection is an arduous but important task. It aims to detect fake news at an early stage before it has widely spread on social networks, when only limited information is available. Early detection is crucial for
fake news, especially due to \textit{validity effect}, which indicates that the more individuals get exposed to certain fake news, the more they may trust it. Meanwhile, it is difficult to correct one's cognition after fake news has gained their trust~\cite{roets2017fake}. Effective early detection of fake news helps take early actions on fake news intervention.
As few temporal information (e.g., the time that users spread the news articles or form relationships) is available in the datasets, the experiment to verify the early detection ability of the proposed approach is designed based on the following intuition.
In our study, each FNN or TNN provides all network information for the corresponding [fake or true] news story. If the dissemination of a news story is at its early stages, the number of spreaders (i.e., nodes) and the involved relationships among spreaders (i.e., edges) should be relatively small compared to when it has been widely spread. Hence, we randomly select a certain proportion ($\in [0,1]$) of nodes or edges for each FNN or TNN and detect fake news on these [sub-] FNNs and [sub-] TNNs. The results are presented in Figure \ref{fig::earlyDetection}. It can be observed from Figure \ref{fig::earlyDetection} that the proposed approach is generally stable with an accuracy and $F_1$ score between $\sim$0.8 ($\sim$0.7) and $\sim$0.9 ($\sim$0.82) by using PolitiFact (BuzzFeed) data, which is friendly to fake news early detection.

\section{Conclusion}
\label{sec::conclusion}
With the rampancy of fake news and the damage it has inflicted on societies, there is a demand for a deep understanding of fake news and effective approaches to detect it. Integrating empirical studies and social psychological theories, our work can deepen the understanding of fake news by investigating its patterns in social networks. These patterns are further exploited and represented at multiple network levels (i.e., node-, ego-, triad-, community- and network-level) to detect fake news in an explainable way. Experiments on two real-world datasets validate the effectiveness of the proposed approach, which can perform relatively well compared to the state-of-the-art. It should be pointed out that compared to content-based models, the proposed approach can hardly detect fake news before it has been propagated on social media, while it can detect fake news with a stable performance by using limited amount of network (propagation) information and a very small number of training news articles. Additionally, by rarely relying on news content, it provides the other perspective to detect fake news which is being robust to the possible manipulation writing styles by malicious entities. Clearly, the proposed approach can be enhanced by introducing more patterns and user attributes that are defined using network information such as network roles~\cite{henderson2012rolx}, and validated on cross-domain and language fake news data to assess its generalization power. Both will be part of our future work.

\balance
\vspace{2mm}
\bibliographystyle{abbrv}
\bibliography{references}

\end{document}